\documentclass[aps,superscriptaddress,10pt,nofootinbib,notitlepage,twocolumn,prd]{revtex4-1}

\usepackage{hyperref}
\usepackage{xcolor}
\usepackage{graphicx}
\usepackage{physics}
\usepackage{bm}
\usepackage{amssymb}

\newcommand{\be}[1]{\begin{equation} #1 \end{equation}}

\graphicspath{{plots/}}

\newcommand{\xf}{\xflash}
\newcommand{\zf}{z_{\rm flash}}

\newcommand{\trise}{t_{\rm rise}}
\newcommand{\zre}{z_{\rm re}}

\newcommand{\SM}[1]{\text{\sc sm}}
\newcommand{\KKLT}[1]{\text{$\mathrm{KKLT}$}}
\newcommand{\LVS}[1]{\text{$\mathrm{LVS}$}}
\def\be{\begin{equation}}
\def\ee{\end{equation}}

\newcommand{\taureio}{\tau} % {\tau_{\rm reio}}

\newcommand{\xflash}{x_{\rm flash}} %{x_\mathrm{Flash}}
\newcommand{\zflash}{z_{\rm flash}} %{x_\mathrm{Flash}}
\begin{document}
% --------------------------------%
%%%%%%%%%%%%%%%%%%%%%%%%%%%%%%%%%%%

\title{Tangled $\tau$: Planck CMB Constraints on Flash Reionization}

\author{Param Upadhyay}
\email{upadhya6@myumanitoba.ca}
\affiliation{Department of Physics, University of Winnipeg, Winnipeg MB, R3B 2E9, Canada}
\affiliation{Department of Physics \& Astronomy, University of Manitoba, Winnipeg, MB R3T 2N2, Canada}

\author{Elisa G. M. Ferreira}
\email{elisa.ferreira@ipmu.jp}
\affiliation{Kavli IPMU (WPI), UTIAS, The University of Tokyo, 5-1-5 Kashiwanoha, Kashiwa, Chiba 277-8583, Japan}
\affiliation{Center for Data-Driven Discovery, Kavli IPMU (WPI), UTIAS, The University of Tokyo, Kashiwa, Chiba 277-8583, Japan}

\author{Evan McDonough}
\email{e.mcdonough@uwinnipeg.ca}
\affiliation{Department of Physics, University of Winnipeg, Winnipeg MB, R3B 2E9, Canada}

\begin{abstract}
Recent cosmological tensions have renewed interest in the role of reionization in parameter inference, and flash reionization has been proposed as a transient high-redshift ionization episode that can increase the Thomson optical depth while remaining compatible with CMB polarization data. In this work, we present the first Markov Chain Monte Carlo analysis of the flash reionization model fit to Planck cosmic microwave background temperature, polarization, and lensing data. We perform this analysis using Planck PR4 data in $\Lambda$CDM with flash reionization.
We find that CMB data allow a wide range for the redshift of the flash, $\zflash$, but with constraints that are tightly correlated with the peak ionization fraction $\xflash$. We find constraints $\xflash=0.24 ^{+0.13} _{-0.07}$ and $\xflash = 0.15 ^{+0.07} _{ - 0.05}$ for the benchmark flash reionization scenarios with fixed $\zflash=20$ and $25$ respectively. The standard $\Lambda$CDM parameters do not exhibit any significant shifts, and in particular, the total optical depth to reionization is not appreciably changed,  with $\tau=0.061 \pm 0.007$ in flash reionization as compared to $\tau=0.059\pm 0.006$ in the standard tanh parametrization. 
We repeat this analysis for Planck PR3 data in place of PR4, and find that the mild preference for a flash is replaced by 95\% CL upper bounds, given by $x_{\rm flash}<0.28$ and $x_{\rm flash}<0.18$ for $z_{\rm flash}=20$ and $25$, respectively, from Planck PR3 data. 
\end{abstract}

\maketitle

%%%%%%%%%%%%%%%%%%%%%%%%%%%%%%%%%%%
%%%%%%%%%%%%%%%%%%%%%%%%%%%%%%%%%%%
%%%%%%%%%%%%%%%%%%%%%%%%%%%%%%%%%%%
%%%%%%%%%%%%%%%%%%%%%%%%%%%%%%%%%%%
\section{Introduction}
\label{sec:intro}
\parskip 5pt

The epoch of reionization contains a multitude of physical phenomena, from cosmic dawn and the first stars to the complete reionization of the intergalactic medium, connecting subatomic physics to the universe on the largest scales. Yet in the broader context of the $\Lambda$CDM cosmological model, reionization is often parametrized as a smooth process described by a single free parameter, namely the optical depth to reionization, or equivalently the redshift of reionization.

The relevance of reionization to tests of $\Lambda$CDM and cosmological parameter inference has gained renewed attention in light of the tension between CMB-inferred parameters and baryon acoustic oscillation measurements from DESI DR2, the so-called BAO-CMB tension \cite{DESIDR2,SPT-3G:2025bzu,Ferreira:2025lrd,Ye:2025ark}. One possible way to reduce this tension is to increase the optical depth to reionization, since a larger $\tau$ changes the CMB-inferred amplitude and shifts the parameters inferred from the CMB in a direction more compatible with BAO measurements \cite{Jhaveri:2025neg,Sailer:2025lxj}. This point is important not only within $\Lambda$CDM~\cite{Ferreira:2025lrd}, but also for the interpretation of extended models, where assumptions about $\tau$ can affect conclusions about neutrino mass, spatial curvature, or dynamical dark energy.

In parallel, observations of high-redshift quasars and massive black holes by HST and JWST have sharpened questions about the formation of the first luminous objects and black-hole seeds \cite{Maiolino:2023zdu,2024Natur.628...57F}. The scenario of flash reionization \cite{Tan2025,TanKomatsu2025,Aggarwal:2026ogm} connects these questions to CMB cosmology. In this picture, Pop.~III.1 stars generate a transient high-redshift ionization episode before the usual late-time reionization transition. This early flash can contribute to the total integrated optical depth and may also be connected to the formation of the supermassive black holes inferred from high-redshift quasars \cite{Petkova:2026clg,Chon2026}.

There is a long history of work on double or non-monotonic reionization histories \cite{Cen:2002zc,Wyithe:2002qu,Holder:2003eb,Naselsky:2003dp,Colombo:2004uh,Furlanetto:2004nt}. More recently, Sierra et al. (Ref.~\cite{Sierra:2026vtz}) discussed the possibility of an alternative reionization scenario from primordial black hole Hawking radiation. The flash reionization scenario considered here is a particular realization of this broader idea, but with several distinctive features: the ionization episode is fast, occurs at comparatively high redshift, and is motivated by a specific Pop.~III.1 astrophysical picture.

From the CMB perspective, the redshift of the scattering is essential. Since scattering at higher redshift projects the reionization-generated polarization to somewhat smaller angular scales, \textit{i.e.}, larger multipoles, a flash contribution does not produce the same low-$\ell$ $EE$ spectrum as an equivalent increase in the duration or redshift of a standard tanh-like reionization history. This is the key reason why flash histories can be tested with the shape of the reionization bump, rather than only through the total optical depth.

Previous studies of flash reionization considered benchmark histories with $z_{\rm flash}\simeq 20 - 25$ and large flash amplitudes. Tan and Komatsu (Ref.~\cite{TanKomatsu2025}) showed that such histories can raise the optical depth while changing the shape of the low-$\ell$ reionization bump, shifting power away from the lowest multipoles and toward somewhat larger low multipoles. Aggarwal et al. (Ref.~\cite{Aggarwal:2026ogm}) further explored whether Pop.~III.1 flash histories can generate a large optical depth while evading Ly$\alpha$ forest and pkSZ constraints, thereby providing an astrophysically motivated realization of the high-$\tau$ scenarios discussed in connection with the BAO-CMB tension.
These results make flash reionization a compelling scenario to test, but fixed benchmark histories cannot establish whether the mechanism survives cosmological parameter inference.  
A full MCMC analysis is therefore required to determine whether flash reionization can genuinely support a high-$\tau$ interpretation and to assess its implications for the BAO-CMB tension.

In this work, we perform the first MCMC analysis of Planck temperature, polarization, and lensing data in the context $\Lambda$CDM with flash reionization. We model the ionization history as the combination of a phenomenological high-redshift flash and the standard late-time tanh reionization transition, and we sample the flash parameters simultaneously with the standard cosmological parameters.

We use Planck PR4 as our benchmark CMB data set and compare with Planck PR3 to assess the dependence of the result on the Planck likelihood choice, in particular the treatment of large-scale polarization. We also compare the inferred constraints in the compressed $(\Omega_m,r_dh)$ plane with DESI DR2 BAO. This allows us to test whether a localized high-redshift ionization episode remains viable after marginalization over the flash parameters, and whether the resulting optical-depth and parameter shifts are large enough to be relevant for the BAO-CMB tension.

Our main result is that Planck PR4 allows a modest flash contribution. The full analysis does not lead to the same benchmark high-$\tau$ histories. In the all-free flash model, the flash redshift and duration are only weakly constrained, and the marginalized total optical depth remains close to the standard tanh result. For fixed benchmark redshifts, $z_{\rm flash}=20$ and $z_{\rm flash}=25$, Planck PR4 constrains the flash amplitudes to values below those assumed in previous fixed-history studies. Thus, the CMB can allow an early ionization contribution, but the evidence for a large increase in the total optical depth is limited once the full parameter space is marginalized over.

Our analysis shows that several important features of flash reionization only become visible in the full parameter inference. Fixed benchmark histories can illustrate how an early ionization episode modifies the low-$\ell$ polarization spectrum, but they cannot determine how the result changes after marginalization over the flash parameters, the late reionization history, and the standard cosmological parameters. We find that Planck PR4 allows a modest flash contribution, but the result must be interpreted together with marginalization, the optical-depth budget, and the prior structure induced by the flash parametrization. In particular, because $\tau$ is derived from the full ionization history rather than sampled directly, the priors on the flash parameters induce a nontrivial prior on the total optical depth. We explicitly test this induced prior and include it in the interpretation of the marginalized constraints. This is distinct from broader prior-volume and marginalization effects, which may also enter when some flash directions are weakly constrained. We therefore present the constraints together with the induced prior on $\tau$ and with the decomposition of the optical depth into early and late contributions.

This paper is organized as follows. In Sec.~\ref{sec:flash} we introduce the phenomenological flash-reionization model and its CMB signatures. In Sec.~\ref{sec:datasets} we describe the data sets, likelihoods, and MCMC methodology. In Sec.~\ref{sec:constraints} we present the Planck constraints, including fixed-redshift analyses, flash-duration dependence, CMB lensing, and the PR3--PR4 comparison. We discuss our results and conclude in Sec.~\ref{sec:discussion}.

%%%%%%%%%%%%%%%%%%%%%%%%%%%%%%%%%%%
%%%%%%%%%%%%%%%%%%%%%%%%%%%%%%%%%%%
%%%%%%%%%%%%%%%%%%%%%%%%%%%%%%%%%%%
%%%%%%%%%%%%%%%%%%%%%%%%%%%%%%%%%%%
\section{Flash Reionization}
\label{sec:flash}
%%%%%%%%%%%%%%%%%%%%%%%%%%%%%%%%%%%
%%%%%%%%%%%%%%%%%%%%%%%%%%%%%%%%%%%
%%%%%%%%%%%%%%%%%%%%%%%%%%%%%%%%%%%
%%%%%%%%%%%%%%%%%%%%%%%%%%%%%%%%%%%

Here we review the flash reionization scenario following Refs.~\cite{Tan2025,TanKomatsu2025,Aggarwal:2026ogm}.
We adopt a phenomenological history, with a transient (``flash'') phase of reionization occurring at early times and smooth reionization transition at late times. The early rise and decay follow the prescription discussed in Refs.~\cite{Tan2025,TanKomatsu2025,Aggarwal:2026ogm}. We define its normalization and the treatment of overlap between the two ionization epochs below.

Let $x_e=n_e/n_{\rm H}$ denote the free-electron abundance per hydrogen nucleus. The sampled flash amplitude $\xf$ is the peak hydrogen contribution before the helium correction; it lies between zero and one. We use the matter-dominated time--redshift relation to define the flash shape,
\begin{equation}
 t_{\rm MD}(z)=\frac{2}{3H_0\sqrt{\Omega_m}}(1+z)^{-3/2}.
 \label{eq:time}
\end{equation}
This approximation specifies the time dependence of the phenomenological ionization history. The background evolution and CMB spectra are otherwise computed with the Boltzmann solver.

Writing $t_f=t_{\rm MD}(\zf)$ and $t_{\rm form}=t_f-\trise$, the hydrogen flash contribution is
\begin{equation}
 x_{\rm H}^{\rm flash}(t)=\xf
 \begin{cases}
 0, &t<t_{\rm form},\\
 (t-t_{\rm form})/\trise, &t_{\rm form}\leq t\leq t_f,\\
 \exp[-(t-t_f)/t_{\rm rec}], &t>t_f.
 \end{cases}
 \label{eq:flash}
\end{equation}
This parametrization describes a linear growth in $x_{\rm H}$, that begins at the time $t_{\rm form}$ when Population III.1 stars form, and lasts for a time $\trise$ at which point $x_{\rm H}$ has reached a peak value of $\xflash$ and subsequently decreases exponentially on a timescale $t_{\rm rec}$ corresponding to the recombination timescale in the ionized regions. Following Refs.~\cite{Tan2025,TanKomatsu2025,Aggarwal:2026ogm}, we model the recombination timescale as,
\begin{equation}
 t_{\rm rec}=\left[\alpha_{\rm rec}\,\delta\,n_{{\rm H},0}
 (1+\zf)^3\right]^{-1},
 \label{eq:trec}
\end{equation}
with $\alpha_{\rm rec}=1.08\times10^{-13}\,\mathrm{cm^3\,s^{-1}}$ and $\delta=3$.

In Refs.~\cite{Tan2025,TanKomatsu2025,Aggarwal:2026ogm}, the flash is further decomposed into a volume fraction of the IGM which is ionized, $f_{i,{\rm vol}}$,  and a peak ionization fraction in those regions $x_{i, {\rm max}}$. As these two parameters are linearly degenerate in their contribution to the optical depth \cite{Tan2025,TanKomatsu2025,Aggarwal:2026ogm}, in this work we combine them into a single parameter $\xflash$.

The late phase of reionization is described by the standard tanh parametrization of reionization \cite{Lewis:2008wr}, which we denote $x_e^{\tanh}(z)$, that is  widely used in cosmological data analysis. In this parametrization the hydrogen and first-helium transition is centered at the sampled $\zre$, with fixed width $\Delta z=0.5$ and exponent $3/2$. The second helium transition is centered at $z=3.5$ with width $0.5$. We include singly ionized helium in $x_e$ using the rescaling factor $f \equiv 1+n_{\rm He}/n_{\rm H} =1.08$ \cite{Aggarwal:2026ogm}. We combine the early and late reionization histories as (see \cite{Aggarwal:2026ogm}):
\begin{equation}
 x_e(z)=\max\left[x_e^{\tanh}(z), x_{e}^{\rm flash}(z)\right].
 \label{eq:combine}
\end{equation}
Equation~\eqref{eq:combine} specifies the overlap prescription, including when the two epochs are not well separated. The zero-amplitude limit recovers the late tanh history;
a sufficiently late or weak flash can easily be hidden beneath the tanh component in Eq.~\eqref{eq:combine}.

\begin{figure}[h!]
    \centering
    \includegraphics[width=\linewidth]{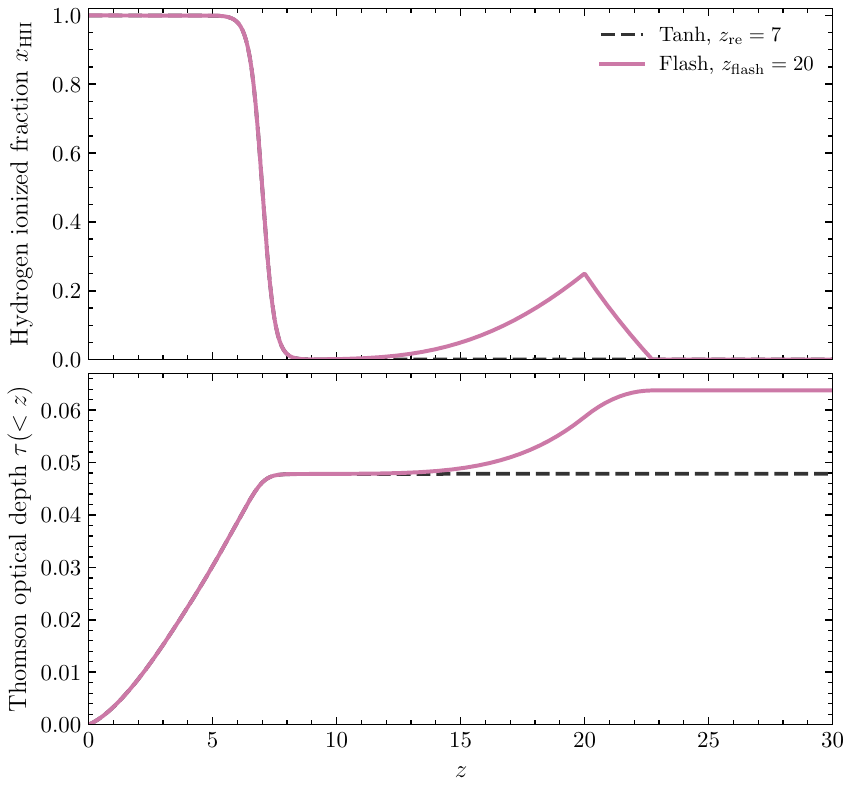}
    \caption{Reionization history and optical depth in the flash reionization scenario. We use the parametrization Eq.~\ref{eq:flash} to model the flash, and prescription Eq.~\ref{eq:combine} to combine the flash and late-time reionization epochs. The optical depth is computed using Eq.~\ref{eq:tau} using a modified version of the \texttt{CLASS} Einstein-Boltzmann solver \cite{CLASS}.}
    \label{fig:flash-history}
\end{figure}

Given the ionization history, one may compute the scattering optical depth over a redshift interval as
\begin{equation}
 \tau(z_1,z_2)=c\sigma_T n_{{\rm H},0}
 \int_{z_1}^{z_2}\frac{(1+z)^2}{H(z)}x_e(z)\,dz,
 \label{eq:tau}
\end{equation}
where $\sigma_T$ is the Thomson cross section. A fiducial example of $\tau(z)\equiv \tau(0,z)$ is shown in Fig.~\ref{fig:flash-history}, where we show the ionization fraction and optical depth for a flash with parameters $z_{\rm flash}=20$, $x_{\rm flash}=0.25$, and $t_{\rm rise}=30$ Myr.

Refs.~\cite{Tan2025,TanKomatsu2025,Aggarwal:2026ogm} provide benchmark examples of flash reionization, with $\zflash=20$ or $25$, $\xflash=0.5$ (corresponding to $f_{i,{\rm vol}}=0.5$ and $x_{i,{\rm max}}=1.0$ in Ref \cite{Aggarwal:2026ogm}), and $t_{\rm rise}=30$ Myr. In our work we will consider  $\zflash$, $\xflash$, and $\trise$ to be free parameters of the flash model, which we vary along with the standard $\Lambda$CDM parameters aside from $\tau$. One may anticipate the degeneracies in such an analysis already from Eq.~\ref{eq:tau}. In particular, since $(1+z)^2/H(z) \propto \sqrt{1+z}$, one may infer that a later flash (smaller $\zflash$) is degenerate with a larger $\xflash$.

The success of this model lies in part in its ability to match cosmic microwave background polarization, namely low-$\ell$ EE polarization power spectrum data, with a larger total $\tau$ than would be inferred from data in the conventional tanh parametrization for reionization. This is relevant in light of the tension between CMB and BAO data that may in part be addressed by a larger $\tau$ \cite{Jhaveri:2025neg,Liu:2025bss,Allali:2025yvp,Sailer:2025lxj}. We emphasize, however,  that the CMB does not respond only to the total integrated optical depth. The redshift distribution
of the scattering also determines the angular scales on which
reionization-generated polarization appears.

\begin{figure}[h!]
    \centering
    \includegraphics[width=\linewidth]{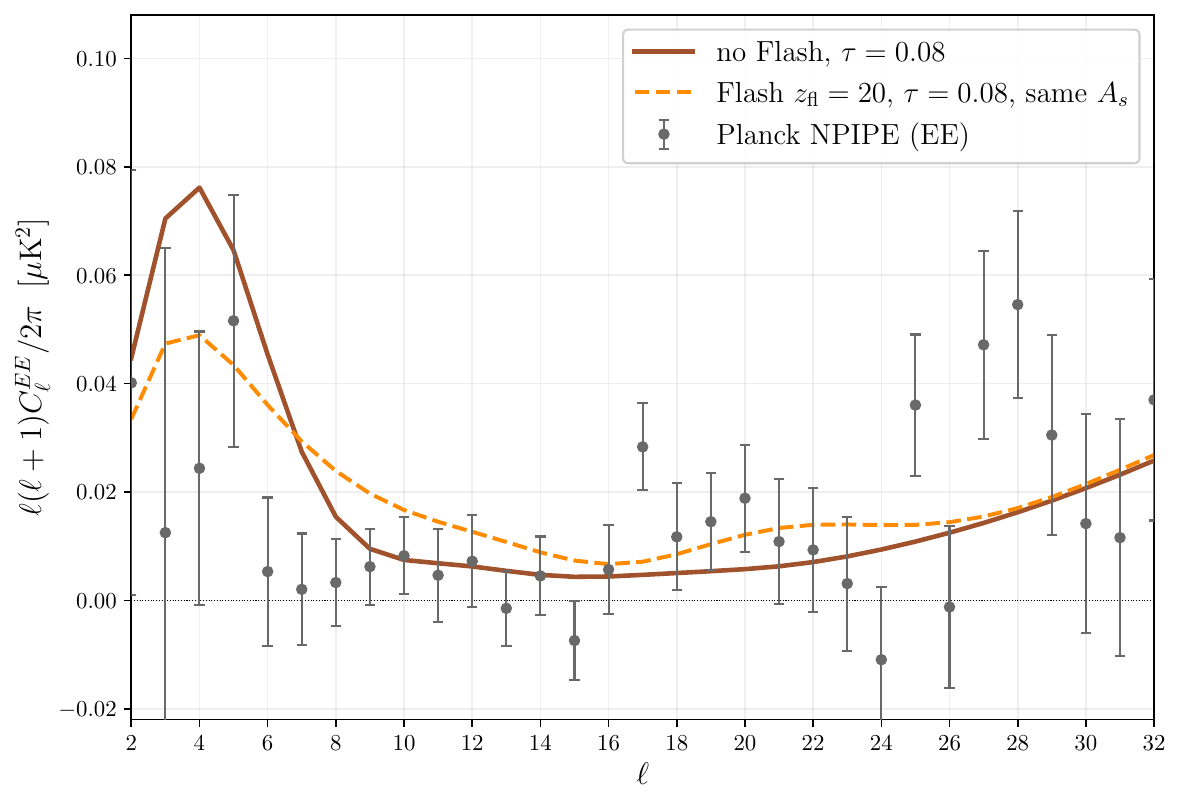}
    \caption{CMB polarization in flash reionization. We show the low-$\ell$ EE power spectrum in a fiducial $\Lambda$CDM cosmology with a standard tanh reionization history (red), and in flash reionization with the same total $\tau$ (orange, dashed). For comparison, we include constraints from Planck PR4 EE power spectrum data.}
    \label{fig:komatsutan}
\end{figure}

This is illustrated in Fig.~\ref{fig:komatsutan}, where we compare a
standard late-reionization history with a flash history having the same
total optical depth, $\tau=0.08$, and the same primordial scalar amplitude
$A_s$. Thomson scattering during reionization generates large-scale
$E$-mode polarization by scattering the local CMB temperature quadrupole.
Scattering at higher redshift occurs when the relevant horizon subtends a
smaller angle on the sky, shifting the associated polarization signal
toward larger multipoles. Consequently, redistributing part of a fixed
optical depth from the usual late-time transition to a high-redshift flash
suppresses the lowest multipoles of the reionization bump while enhancing
power at somewhat larger multipoles.

For the $\zf=20$ example in Fig.~\ref{fig:komatsutan}, the flash model
therefore produces less power at $\ell\lesssim 8$ and more power at
$\ell\gtrsim 8$ than the no-flash model, despite the two models having
identical $\tau$. This reproduces the characteristic behavior identified
in Ref.~\cite{TanKomatsu2025}.

An additional subtlety in analyses of flash reionization, and more generally in analyses of flexible reionization histories, is that a prior distribution in the sampled model parameters is not necessarily preserved when mapped onto a derived quantity such as the optical depth. In the flash model, $\tau$ is not sampled directly, but is computed from the full ionization history. Uniform priors on the flash parameters can therefore induce a non-uniform effective, \textit{implicit prior} on the total $\tau$. This issue has been emphasized in the context of flexible reionization reconstructions and reionization-prior effects \cite{MilleaBouchet2018,Ilic:2025idl,Wang:2026bsq}.

\begin{figure}[h!]
    \includegraphics[width=\linewidth]{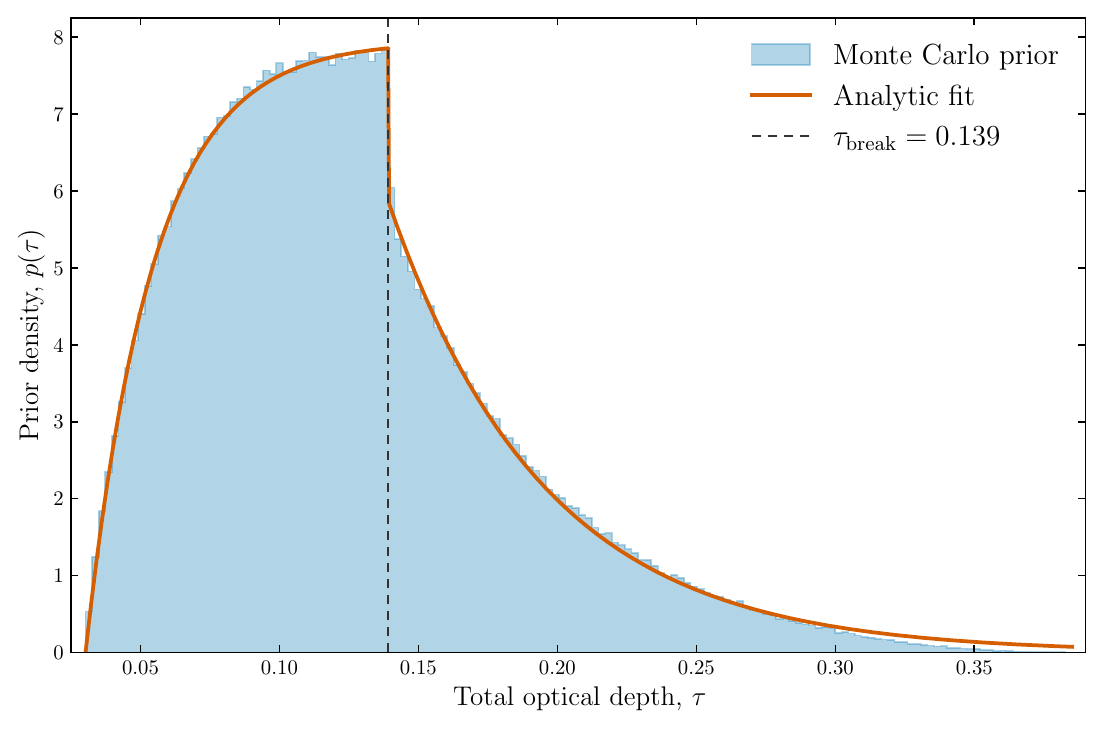}
    \caption{
    Effective prior on $\tau$ from a uniform distribution of the flash parameters in the ranges $z_{\rm flash} = [5,40]$, $\xflash = [0,1]$, with fixed $t_{\rm rise}=30$ Myr, and a uniform prior on the redshift to late time reionization $z_{\rm reio}=[5, 15]$, with other cosmological parameters fixed to  a fiducial $\Lambda$CDM cosmology,
    \label{fig:priorvolume}
    }
\end{figure}

To quantify this effect in the present parametrization, we uniformly sample the flash parameters at fixed $\Lambda$CDM cosmology and fixed $t_{\rm rise}=30\,{\rm Myr}$, and construct the resulting probability distribution of the total integrated optical depth $\tau$. This effective prior is shown in Fig.~\ref{fig:priorvolume}, where we sample $z_{\rm flash}\in[5,40]$ and $x_{\rm flash}\in[0,1]$, and include late time reionization with a uniform prior $z_{\rm reio} \in[5, 15]$. We then bin the resulting $\tau$ values and from this construct the effective prior distribution.

The resulting distribution for the induced prior is not flat in $\tau$, but it has broad support over the range of optical depths relevant for both the standard Planck optical-depth constraint and the higher-$\tau$ values discussed in the flash scenarios considered here. In particular, there is substantial prior volume for $\tau>0.06$, and the induced implicit prior rises monotonically over the range $\tau<0.10$. Conversely, the small-$\tau$ region, $\tau\lesssim0.06$, occupies a relatively small fraction of the prior volume. Thus, the prior does not exclude higher-$\tau$ histories a priori. At the same time, the non-uniform mapping from the flash parameters to $\tau$ means that different ranges of $\tau$ are represented by different amounts of prior volume in the sampled flash-parameter space, and therefore receive different prior weight in the induced distribution. This should be taken into account when interpreting marginalized Bayesian constraints on the optical depth. We therefore use Fig.~\ref{fig:priorvolume} as a diagnostic when interpreting the Bayesian constraints on $\tau$ below.

%%%%%%%%%%%%%%%%%%%%%%%%%%%%%%%%%%%
%%%%%%%%%%%%%%%%%%%%%%%%%%%%%%%%%%%
%%%%%%%%%%%%%%%%%%%%%%%%%%%%%%%%%%%
%%%%%%%%%%%%%%%%%%%%%%%%%%%%%%%%%%%
\section{Data sets and methodology}
\label{sec:datasets}
%%%%%%%%%%%%%%%%%%%%%%%%%%%%%%%%%%%
%%%%%%%%%%%%%%%%%%%%%%%%%%%%%%%%%%%
%%%%%%%%%%%%%%%%%%%%%%%%%%%%%%%%%%%
%%%%%%%%%%%%%%%%%%%%%%%%%%%%%%%%%%%

The focus of this work is Planck data
from PR3 and PR4. We make use of the following likelihoods:
\begin{itemize}
\item Low-$\ell$ TT:  \texttt{Commander} likelihood for low-$\ell$ ($\ell < 30$) temperature anisotropy data from Planck PR3~\cite{Planck2018Likelihood}.
\item Low-$\ell$ EE: We consider the SimAll EE likelihood based on Planck PR3 data~\cite{Planck2018Likelihood} and the \texttt{LoLLiPoP} low-$\ell$ polarization likelihood based on Planck PR4 data~\cite{planck20-57,Tristram2024}. 
\item High-$\ell$ TT/EE/TE: We consider the Plik likelihood based on PR3 data~\cite{Planck2018Likelihood}  and \texttt{HiLLiPoP}  high-$\ell$ temperature and polarization likelihood based on PR4 data~\cite{planck20-57,Tristram2024}.
\item CMB lensing: CMB lensing potential $\phi\phi$ power spectrum likelihood from Planck PR3 \cite{Planck:2018lbu} and Planck PR4 \cite{Carron:2022eyg} data.
\end{itemize}
We organize these into two data set combinations:
\begin{itemize}
    \item {\bf Planck PR4}: Planck 2018 commander low-$\ell$ TT, and Planck 2020 NPIPE LoLLiPoP low-$\ell$ polarization and HiLLiPoP high-$\ell$ TTTEEE, combined with Planck PR4 lensing.
    \item {\bf Planck PR3}: Planck 2018 commander low-$\ell$ TT, SimAll low-ell polarization, Plik PR3 high-ell, and PR3 lensing.
\end{itemize}
We will also compare these constraints with DESI DR2 BAO data \cite{DESIDR2}. In the context of $\Lambda$CDM, and independent of the details of reionization, the BAO data can be compressed into the parameters $r_d h$ and $\Omega_M$, as discussed in e.g.~\cite{McDonough:2025lzo,Ferreira:2025lrd}. The DESI DR2 constraints on these parameters are given by $r_dh=101.54\pm0.73$ and $\Omega_m=0.2975\pm 0.0086$. These constraints are in modest tension with CMB inferences of these parameters, which has been referred to as the BAO-CMB tension \cite{SPT-3G:2025bzu,McDonough:2025lzo}.

We perform Markov Chain Monte Carlo (MCMC) analyses using \texttt{Cobaya} \cite{torrado_lewis_2019}\footnote{\url{https://github.com/CobayaSampler/cobaya/tree/master}} and a modified version of the Einstein-Boltzmann solver \texttt{CLASS} \cite{Diego_Blas_2011}. We enforce a convergence criterion on MCMC chains of Gelman-Rubin statistic $R-1 <0.05$ and note that most of our analyses have $R-1$ much smaller than this. We use \texttt{GetDist} ~\cite{GetDist}\footnote{\url{https://github.com/cmbant/getdist}} to plot the results.

%%%%%%%%%%%%%%%%%%%%%%%%%%%%%%%%%%%%%%%%%%%%%%%%%%%%%%
%%%%%%%%%%%%%%%%%%%%%%%%%%%%%%%%%%%%%%%%%%%%%%%%%%%%%%
%%%%%%%%%%%%%%%%%%%%%%%%%%%%%%%%%%%%%%%%%%%%%%%%%%%%%%
%%%%%%%%%%%%%%%%%%%%%%%%%%%%%%%%%%%%%%%%%%%%%%%%%%%%%%
\section{Constraints from Planck CMB Data}
\label{sec:constraints}

We perform MCMC analyses of $\Lambda$CDM with flash reionization fit to data sets described above, and compare with constraints from the standard tanh parametrization. In what follows, $\tau$ will refer to the total integrated optical depth, including both the flash and tanh contributions to reionization.

%%%%%%%%%%%%%%%%%%%%%%%%%%%%%%%%%%%%%%%%%%%%%%%%%%%%%%
%%%%%%%%%%%%%%%%%%%%%%%%%%%%%%%%%%%%%%%%%%%%%%%%%%%%%%
\subsection{Constraints from Planck PR4}
%%%%%%%%%%%%%%%%%%%%%%%%%%%%%%%%%%%%%%%%%%%%%%%%%%%%%%
%%%%%%%%%%%%%%%%%%%%%%%%%%%%%%%%%%%%%%%%%%%%%%%%%%%%%%

\begin{figure*}
    \centering
    \includegraphics[width=\linewidth]{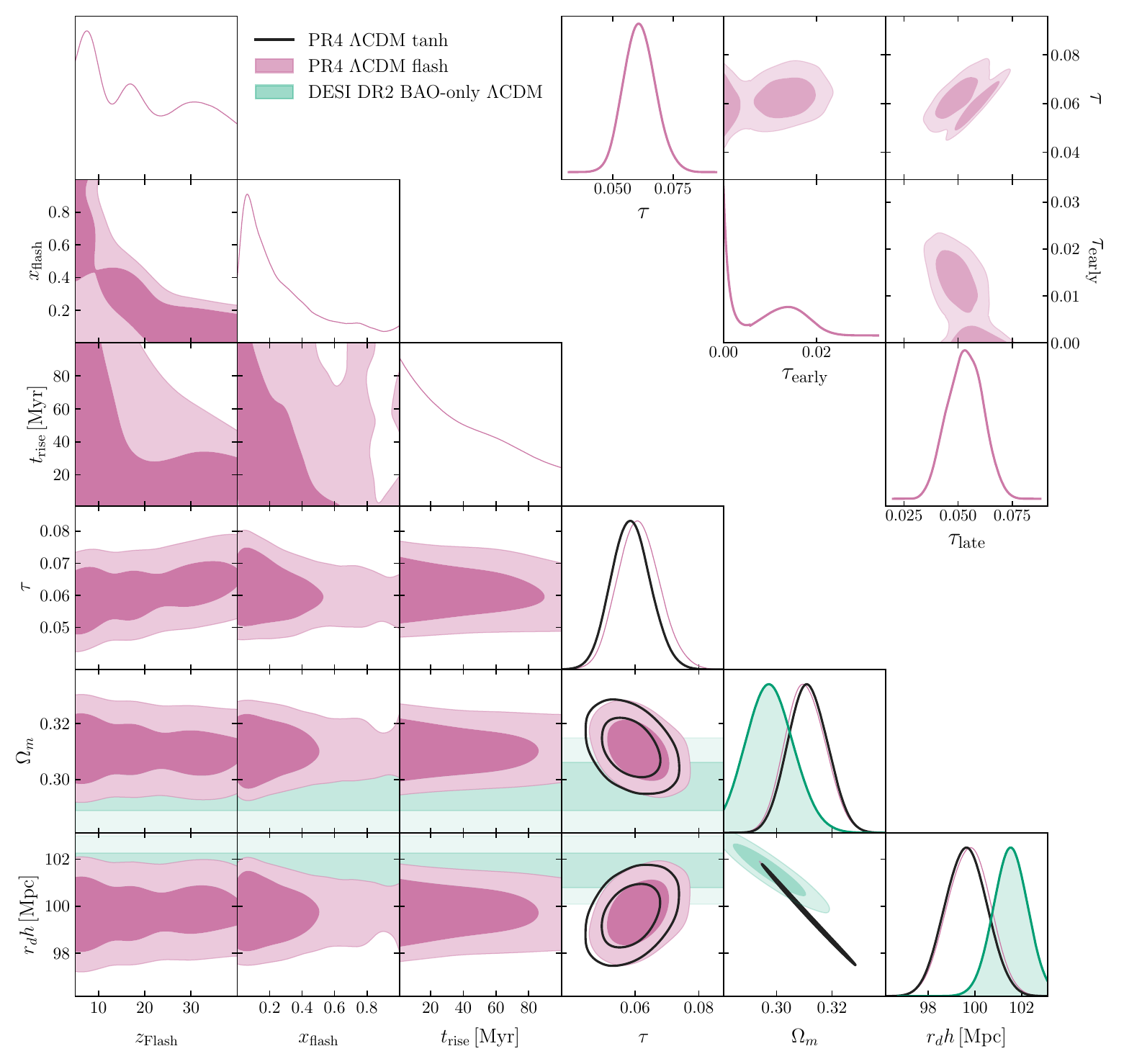}
    
    \caption{  Constraints on $\Lambda$CDM with flash reionization in the fit to CMB data from Planck PR4. We vary the redshift of the flash $z_{\rm flash}$, the peak ionization fraction $\xflash$, and the duration of the onset of the flash $t_{\rm rise}$, along with the standard $\Lambda$CDM parameters aside from $\tau$. Here, $\tau_{\rm early}$ is integrated between $15 < z < 80$ and $\tau_{\rm late}$ is $0 < z < 15$. We include for comparison the constraints in the context of the baseline $\Lambda$CDM tanh parametrization of reionization, and the constraints from DESI DR2 BAO. }
    \label{fig:flagship}
\end{figure*}

We begin with constraints from Planck PR4 temperature, polarization, and lensing data as described in Sec.~\ref{sec:datasets}. We adopt broad uniform priors on flash reionization parameters $z_{\rm flash}$,  $\xflash$, and $t_{\rm rise}$, and on the standard $\Lambda$CDM parameters aside from $\tau$, which is a derived parameter in the flash model, computed from Eq.~\ref{eq:tau}. Concretely we adopt uniform priors in the ranges  $z_{\rm flash}=[5 ,40]$, $\xflash=[0,1]$, and $t_{\rm rise}=[1,100]$ Myr. The results are shown in Fig.~\ref{fig:flagship}, Tab.~\ref{tab:newFlagship}, and Tab.~\ref{table:chi2_flagship}, for the marginalized posterior distributions, parameter constraints, and $\chi^2$-statistics of the best-fit model, respectively. For comparison we include constraints on $\Lambda$CDM with the standard tanh parametrization of reionization with $\taureio$ a sampled parameter, and include constraints on $r_d h$ and $\Omega_m$ from DESI DR2 BAO. A supplementary of parameter constraints table is given in Tab.~\ref{tab:theta_s_flagship_full}.

\begin{table}[h!]
\centering
Constraints from Planck PR4\\
\begin{tabular}{|l|c|c|}
\hline\hline
Parameter & tanh & Flash \\
\hline\hline
{$z_\mathrm{flash}$} & $-$ & $< 37.4\;(34.7)$ \\
$\xflash$ & $-$ & $< 0.83\;(0.14)$ \\
{$t_\mathrm{rise}\,[\mathrm{Myr}]$} & $-$ & $< 90.5\;(9.5)$ \\
{$\tau_\mathrm{reio}$} & $0.059\;(0.059)\pm 0.006$ & $0.061\;(0.067)\pm 0.007$ \\
$\Omega_\mathrm{m}$ & $0.311\;(0.311)\pm 0.007$ & $0.310\;(0.307)\pm 0.007$ \\
$r_\mathrm{d}h\,[\mathrm{Mpc}]$ & $99.6\;(99.7)\pm 0.9$ & $99.7\;(100.1)\pm 0.9$ \\
\hline
\end{tabular}
\caption{Constraints shown in Fig.~\ref{fig:flagship} (see Tab.~\ref{tab:theta_s_flagship_full} for full parameter constraints). The flash model uses Planck PR4 with $z_\mathrm{flash}$, $\xflash$, and $t_\mathrm{rise}$ sampled jointly, along with the $\Lambda$CDM parameters aside from $\taureio$. For comparison we include constraints in the tanh reionization model. Parentheses give the best sampled posterior point, not a dedicated minimizer best fit; one-sided bounds are 95\%.}
\label{tab:newFlagship}
\end{table}

\begin{table}[h!]
\centering
$\chi^2$ statistics from the best-fit $\Lambda$CDM cosmology with tanh and flash reionization models.
\begin{tabular}{lcc}
\hline\hline
Datasets & tanh & Flash \\
\hline
Planck 2018 low-$\ell$ TT
& 22.64 & 22.29 \\

LoLLiPoP low-$\ell$ EE
& 32.67 & 28.66 \\

HiLLiPoP high-$\ell$ TT+TE+EE
& 30505.17 & 30506.16 \\

PR4 lensing
& 9.15 & 9.12 \\

Total $\chi^2$
& 30569.64 & 30566.24 \\

$\Delta\chi^2$
& -- & -3.40 \\
\hline
\end{tabular}
\caption{Planck PR4 temperature, polarization, and lensing $\chi^2$ contributions evaluated at the minimum-$\chi^2$ sampled point for the tanh and flash models. The total includes the low-$\ell$ TT, LoLLiPoP low-$\ell$ EE, HiLLiPoP high-$\ell$ TT+TE+EE, and PR4 lensing likelihoods. We define $\Delta\chi^2 \equiv \chi^2_{\rm Flash}-\chi^2_{\rm tanh}$, such that negative values indicate an improved fit for the flash model.}
\label{table:chi2_flagship}
\end{table}

Fig.~\ref{fig:flagship}, purple contours, shows the constraints on flash reionization. We find the data only weakly constrains both the timing of the flash, $z_{\rm flash}$ and the duration of the flash $t_{\rm rise}$. The 1d posterior for the peak ionization fraction, $\xflash$, is peaked at $x_{e} \sim 0.2$ and exhibits a tail towards larger values, including unity, $\xflash=1$. 

\begin{figure}[h!]
    \centering
    \includegraphics[width=\linewidth]{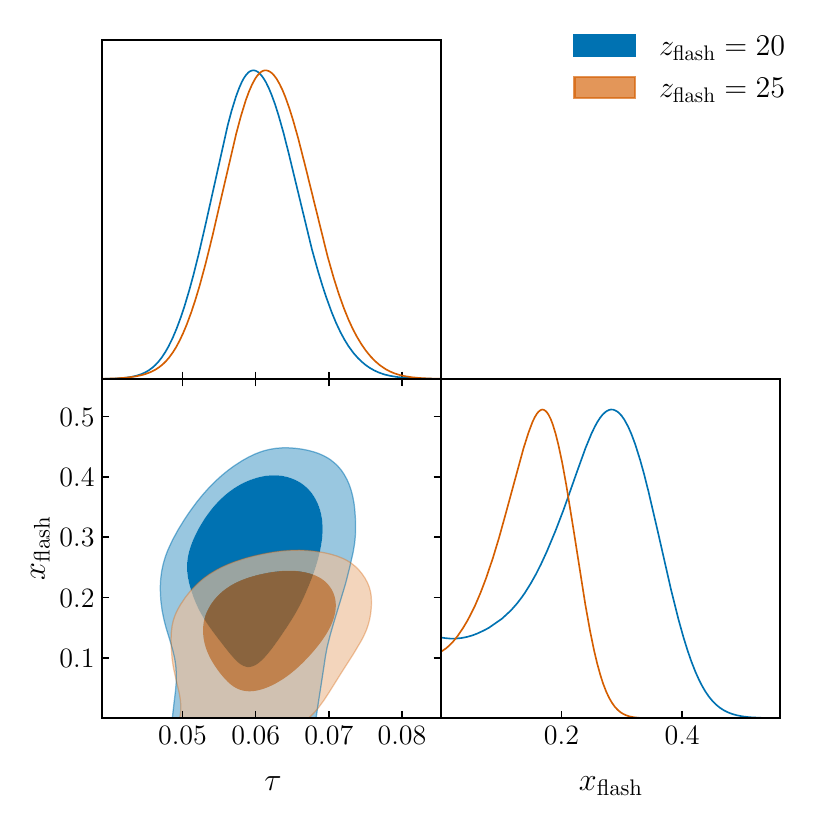}
    \caption{ Constraints at fixed $z_{\rm flash}=20$
(blue) and $25$ (orange) from Planck PR4 data. We fix $t_{\rm rise}=30$~Myr motivated the benchmark models of Ref.~\cite{Tan:2025obi}. The constraint on $\taureio$ is nearly
identical between the two, while $\xflash$ is lower for the
higher-redshift flash.
    \label{fig:fixed}
    }
\end{figure}

However, the constraints on $\xflash$ and $\zflash$ are tightly correlated. We find that a flash at relatively late times, $z_{\rm flash}<10$, is largely unconstrained in ionization fraction $\xflash$ and duration $t_{\rm rise}$, whereas an early flash, $z_{\rm flash}>20$, in the range advocated for in Refs.\cite{Tan:2025obi,Tan:2025cua,Aggarwal:2026ogm}, is relatively well constrained. To dive deeper into this correlation structure, in Fig.~\ref{fig:fixed} and Tab.~\ref{tab:fixed} we repeat this analysis for fixed values of $z_{\rm flash}=20$ and $z_{\rm flash}=25$, corresponding to the benchmark examples of Refs.~\cite{TanKomatsu2025}.

We find constraints on $\xflash$ given by $\xflash=0.24^{+0.13}_{-0.07}$ and
$\xflash=0.15^{+0.07}_{-0.05}$ for $z_{\rm flash}=20$ and $z_{\rm flash}=25$, respectively. While exhibiting a modest preference for nonzero $\xflash$,
these constraints are below the benchmark amplitude $\xflash=0.5$ used in Refs.~\cite{TanKomatsu2025,Aggarwal:2026ogm}, corresponding to $f_{i,\rm vol}=0.5$ and $x_{i,\max}=1.0$ in the notation of Ref.~\cite{Aggarwal:2026ogm}.
Similarly, we find $\tau=0.060^{+0.005}_{-0.006}$ and $\tau=0.062\pm0.006$ in the two cases, respectively. Thus, in these fixed-redshift benchmark slices, the marginalized total optical depth remains close to the standard tanh result
rather than reaching the higher values, $\tau\simeq0.09$, discussed in connection with the BAO-CMB tension \cite{Aggarwal:2026ogm}.

\begin{table}[h!]
\centering
Constraints from Planck PR4 for fixed $z_{\rm flash}$
\begin{tabular}{|l|c|c|}
\hline\hline
Parameter & $z_\mathrm{flash}=20$ & $z_\mathrm{flash}=25$ \\
\hline\hline
{$\tau_\mathrm{reio}$} & $0.060\;(0.064)^{+0.005}_{-0.006}$ & $0.062\;(0.062)\pm 0.006$ \\
{$z_\mathrm{re}$} & $6.7\;(7.3)^{+0.7}_{-1.0}$ & $7.1\;(7.1)\pm 0.8$ \\
{$z_\mathrm{flash}$} & $20\;\mathrm{(fixed)}$ & $25\;\mathrm{(fixed)}$ \\
{$x_\mathrm{flash}$} & $0.24\;(0.21)^{+0.13}_{-0.07}$ & $0.15\;(0.15)^{+0.07}_{-0.05}$ \\
{$t_\mathrm{rise}\,[\mathrm{Myr}]$} & $30\;\mathrm{(fixed)}$ & $30\;\mathrm{(fixed)}$ \\
$\Omega_\mathrm{m}$ & $0.311\;(0.305)\pm 0.007$ & $0.310\;(0.310)\pm 0.007$ \\
$r_\mathrm{d}h\,[\mathrm{Mpc}]$ & $99.7\;(100.4)\pm 0.9$ & $99.8\;(99.8)\pm 0.9$ \\
\hline
\end{tabular}
\caption{Constraints from Planck PR4 TT+TE+EE, low-$\ell$ TT/EE, and lensing, for flash reionization with $z_{\rm flash}$ fixed to 20 and 25 in the second and third columns, respectively. We fix $t_{\rm rise}=30$ Myr motivated by the benchmark models given in Ref.~\cite{Tan:2025obi}.}
\label{tab:fixed}
\end{table}

\begin{figure}[h!]
    \centering
    \includegraphics[width=\linewidth]{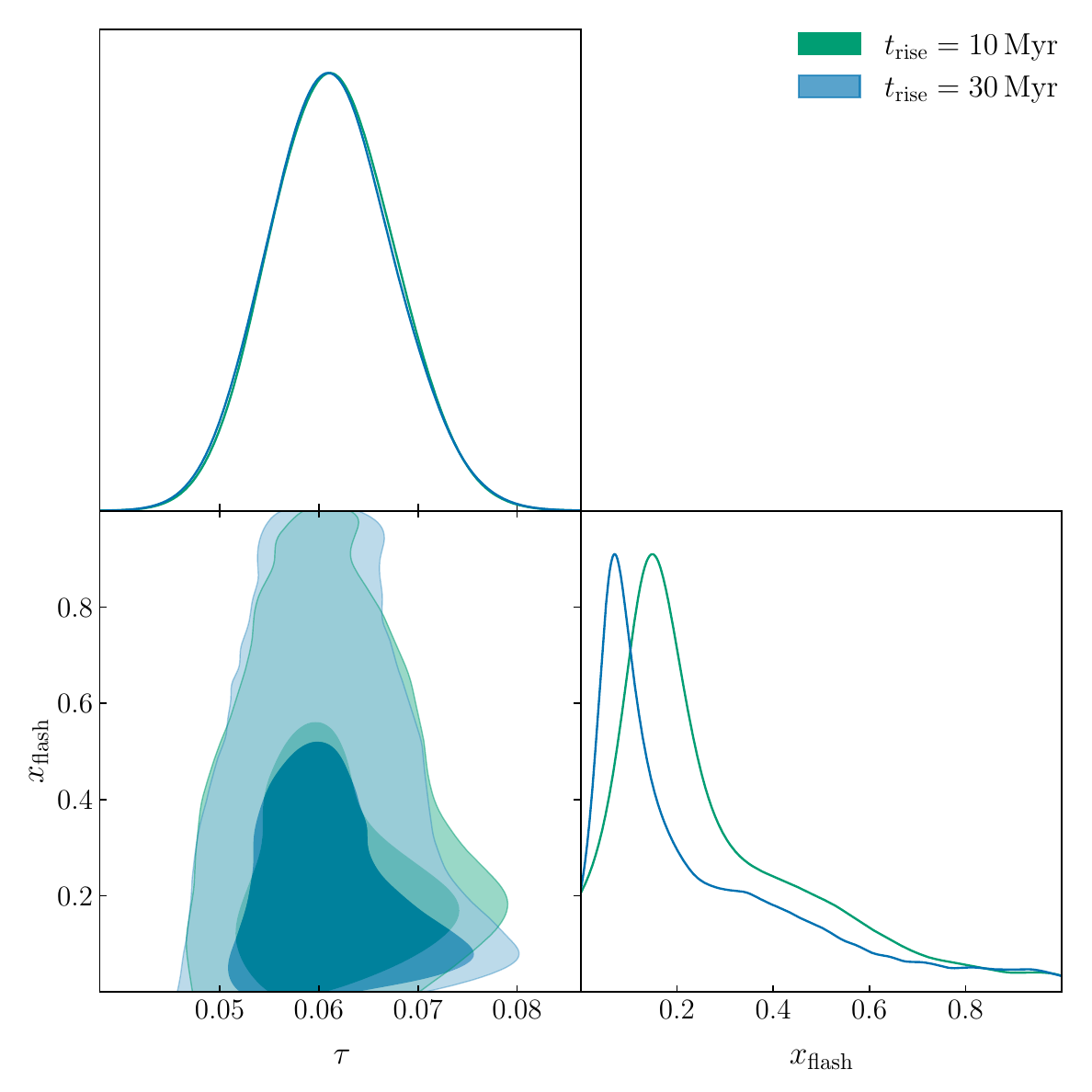}
    \caption{ Constraints at fixed flash rise times $\trise=10$ and $30$ Myr from Planck  PR4 data. Increasing the rise time shifts the peak flash amplitude distribution toward smaller values, while total optical depth and background parameters change comparatively little. A longer ramp can supply a similar scattering contribution with a smaller peak fraction. We find that $\tau$ is completely insensitive to the change in $t_{\rm rise}$.}
    \label{fig:fixedtrise}
\end{figure}

A similar exercise can be performed for $t_{\rm rise}$. In Fig.~\ref{fig:fixedtrise} we show constraints at various fixed values of $t_{\rm rise}$. From this one may appreciate that a shorter flash can accommodate a larger ionization fraction, consistent with the $t_{\rm rise}-z_{\rm flash}$ shown in Fig.~\ref{fig:flagship}, but we note that the total optical depth $\tau$ is insensitive to changes in $\trise$. Parameter constraints are given in Tab.~\ref{tab:varied_fixed_trise}.

We now turn our focus to constraints on $\tau$. Returning to the constraints with all flash parameters varied, Fig.~\ref{fig:flagship}, we find that the constraint on the total combined  $\tau$, including both the flash and late-time reionization contributions, is only marginally shifted from the standard tanh parametrization: we find $\tau=0.061 \pm 0.007$ in flash vs. $\tau=0.059\pm0.006$ in tanh parametrization. Similarly, we find no noticeable shift in the BAO parameters $\Omega_m$ and $r_d h$.

The relative contributions to $\tau$ from the flash and standard late-time reionization are shown in the upper right triangle of Fig.~\ref{fig:flagship}, where we isolate the contribution to Eq.~\eqref{eq:tau} coming from $z<15$ (``$\tau_{\rm late}$'') and $z>15$ (``$\tau_{\rm early}$''). We find that the posterior for $\tau_{\rm early}$ shows significant support at $\tau_{\rm early}=0$ and modest support for a second mode at $\tau_{\rm early}\sim 0.012$. The late contribution is well constrained to $\tau_{\rm late}\sim 0.05$.

We note that these marginalized parameter constraints are offset from the {\it best-fit} flash model, which has $z_{\rm flash}=34.7$ and $\xflash=0.14$, leading to $\taureio=0.067$, as compared to $\taureio=0.061$ in the tanh parametrization. This best-fit model differs significantly from the benchmark examples of~\cite{TanKomatsu2025,Aggarwal:2026ogm}, which have $z_{\rm flash}=20$ or $25$,  $\xflash=0.5$ and $t_{\rm rise}=30\,{\rm Myr}$ in our notation. As shown
in Tab.~\ref{table:chi2_flagship}, the flash model improves the best-fit value mainly through the low-$\ell$ EE likelihood, with smaller changes in the
high-$\ell$ TT, TE, and EE likelihoods. The best-sampled point shows that flash histories with somewhat larger optical depth, $\tau\simeq0.07$, can be accommodated by Planck, while the marginalized posterior remains close to the standard tanh result.

This result should, however, be interpreted carefully. To fully understand it, we need to take into consideration the structure of the full posterior. As we see in Fig.~\ref{fig:flagship}, several directions in the flash parameter space,
in particular $z_{\rm flash}$ and $t_{\rm rise}$, are only weakly constrained or unconstrained by the data. In such a situation, the marginalized posterior can be
sensitive both to these unconstrained regions of parameter space and to how much prior volume they occupy. This prior volume can arise either from the priors placed directly on the sampled parameters or from induced priors on derived quantities, such as the implicit prior on $\tau$.
This can lead to shifts in the inferred parameters through prior-volume/marginalization effects. 
Indeed, the implicit prior shown in Fig.~\ref{fig:priorvolume} has substantial support for $\tau>0.06$ and, thus, higher-$\tau$ histories are not excluded by the implicit prior. At the same time, the all-free marginalized posterior remains close to the standard tanh result. This suggests that the preference for $\tau\simeq0.06$ is not simply a consequence of the one-dimensional implicit prior on $\tau$. A complete separation of likelihood effects, prior-volume effects, and parametrization dependence would require additional analyses.

\begin{figure}[h!]
    \centering
    \includegraphics[width=\linewidth]{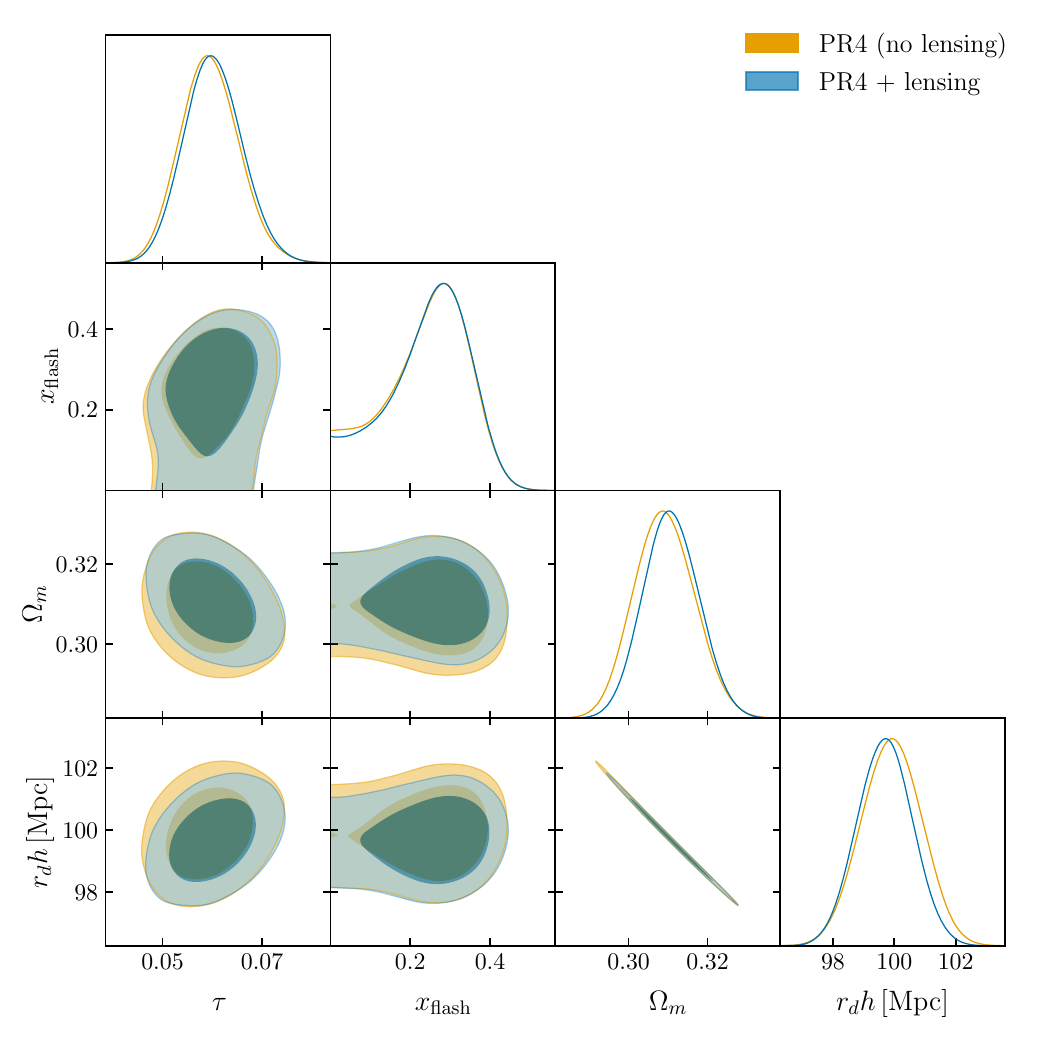}
    \caption{ Flash constraints at $z_{\rm flash}=20$, $t_{\rm rise}=30$~Myr,
with (blue) and without (orange) PR4 lensing. $\taureio$ and
$\xflash$ are essentially unchanged, indicating the flash amplitude
is set by the CMB temperature and polarization spectra rather than CMB lensing data. }
    \label{fig:lensing}
\end{figure}

Finally, we examine the role that CMB lensing plays in constraining flash reionization. In Fig.~\ref{fig:lensing}, we show constraints from PR4 data with and without lensing data. To save computational expense we fix $z_{\rm flash}=20$, $t_{\rm rise}=30$~Myr, and vary $\xflash$ and the standard $\Lambda$CDM parameters aside from $\tau$.  We find that the constraint on $\xflash$ is unchanged, while $\tau$, $\Omega_m$, and $r_dh$ shift slightly, mirroring the impact of CMB lensing on these parameters in the standard tanh model for reionization \cite{McDonough:2025lzo}.

%%%%%%%%%%%%%%%%%%%%%%%%%%%%%%%%%%%%%%%%%%%%%%%%%%%%%%
%%%%%%%%%%%%%%%%%%%%%%%%%%%%%%%%%%%%%%%%%%%%%%%%%%%%%%
\subsection{Comparison of PR3 and PR4}
%%%%%%%%%%%%%%%%%%%%%%%%%%%%%%%%%%%%%%%%%%%%%%%%%%%%%%
%%%%%%%%%%%%%%%%%%%%%%%%%%%%%%%%%%%%%%%%%%%%%%%%%%%%%%

\begin{figure*}
    \centering
    \includegraphics[width=0.8\linewidth]{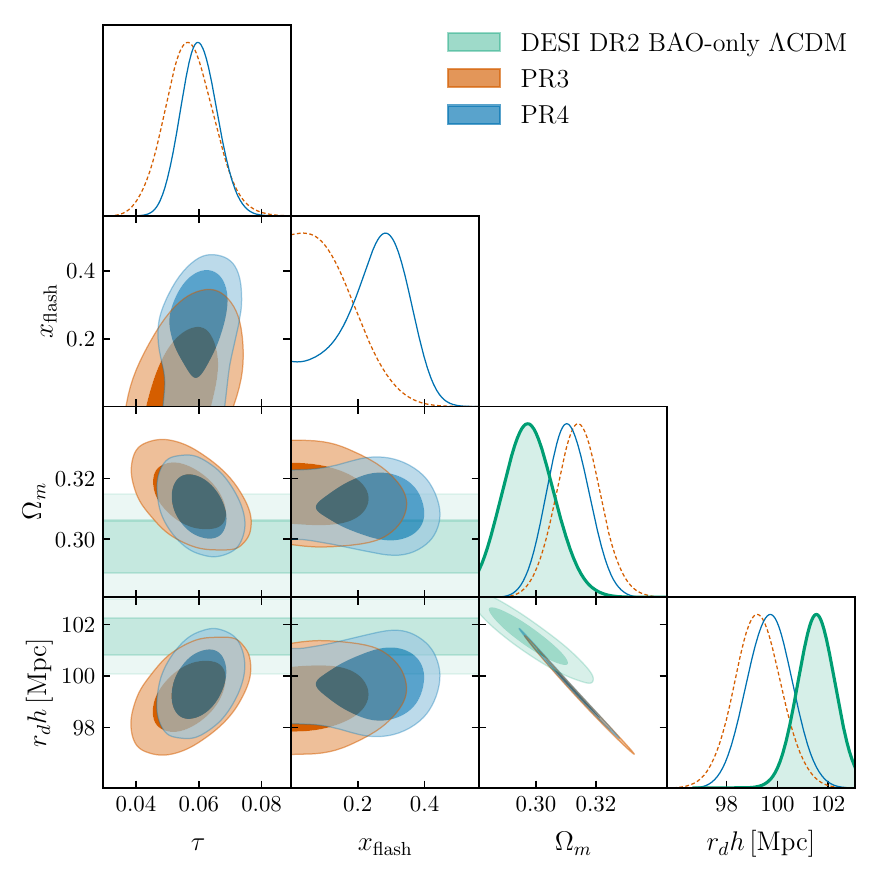}
    \caption{Constraints on $\Lambda$CDM with flash reionization at fixed $z_{\rm flash}=20$ and $t_{\rm rise}=30$~Myr, from Planck
PR3 (orange) and PR4 (blue). PR4 prefers a nonzero
$\xflash$ where PR3 gives only an upper limit, with a correspondingly
higher $\taureio$, lower $\Omega_m$, and larger $r_d h$ mirroring the parameter shifts in the context of the standard tanh parametrization.} 
    \label{fig:PR3PR4}
\end{figure*}

It is important to understand the dependence of the results on the Planck data set and likelihood choice. This is particularly relevant for flash reionization, since the flash contribution is mainly tested through large-scale polarization, which is also the part of the CMB data most directly sensitive to the reionization optical depth. We therefore repeat the flash analysis separately with the Planck PR3 and PR4 data sets. For this comparison we fix $z_{\rm flash}=20$ and $t_{\rm rise}=30\,{\rm Myr}$, corresponding to the lower-redshift benchmark flash model. This gives a controlled comparison of the two Planck likelihoods without introducing the additional degeneracy with $z_{\rm flash}$. The comparison is shown in Fig.~\ref{fig:PR3PR4} and Table~\ref{tab:pr3_vs_pr4_flash_mini}.

In the PR3 analysis we use the SimAll low-$\ell$ polarization likelihood, while in the PR4 analysis we use the NPIPE/LoLLiPoP low-$\ell$ polarization likelihood. The difference between the two results should therefore be understood as a difference in the large-scale polarization constraint on reionization. We do not attempt to isolate which individual multipoles or likelihood ingredients are responsible for this shift. Instead, we use the PR3-PR4 comparison as an empirical check of how sensitive the flash constraints are to the low-$\ell$ polarization data set. The importance of the low-$\ell$ polarization data set for non-standard reionization constraints has already been pointed out in the literature \cite{Ilic:2025idl,TanKomatsu2025}.

\begin{table}[h!]
\centering
Comparison of constraints from Planck PR3 and PR4
\begin{tabular}{|l|c|c|}
\hline\hline
Parameter & PR3 & PR4 \\
\hline\hline
{$\tau_\mathrm{reio}$} & $0.057\;(0.053)^{+0.007}_{-0.008}$ & $0.060\;(0.064)^{+0.005}_{-0.006}$ \\
{$z_\mathrm{re}$} & $7.1\;(7.0)\pm 0.8$ & $6.7\;(7.3)^{+0.7}_{-1.0}$ \\
{$z_\mathrm{flash}$} & $20\;\mathrm{(fixed)}$ & $20\;\mathrm{(fixed)}$ \\
{$x_\mathrm{flash}$} & $< 0.28\;(0.08)$ & $0.24\;(0.21)^{+0.13}_{-0.07}$ \\
{$t_\mathrm{rise}\,[\mathrm{Myr}]$} & $30\;\mathrm{(fixed)}$ & $30\;\mathrm{(fixed)}$ \\
$\Omega_\mathrm{m}$ & $0.314\;(0.314)\pm 0.007$ & $0.311\;(0.305)\pm 0.007$ \\
$r_\mathrm{d}h\,[\mathrm{Mpc}]$ & $99.2\;(99.2)\pm 0.9$ & $99.7\;(100.4)\pm 0.9$ \\
\hline
\end{tabular}
\caption{Planck PR3 versus PR4 flash constraints at $z_\mathrm{flash}=20$ and $t_\mathrm{rise}=30\,\mathrm{Myr}$. Parentheses give the best sampled posterior point, not a dedicated minimizer best fit; one-sided bounds are 95\%.}
\label{tab:pr3_vs_pr4_flash_mini}
\end{table}

We find that PR4 allows a larger flash contribution than PR3. In the standard tanh model, the inferred optical depth shifts from $\taureio=0.054\pm0.007$ with PR3 to $\taureio=0.059\pm0.006$ with PR4 (see Tab.~\ref{tab:pr3_vs_pr4_lcdm_and_flash_big}). The same trend appears in the flash model. For fixed $z_{\rm flash}=20$ and $t_{\rm rise}=30\,{\rm Myr}$, PR3 gives $\taureio=0.057^{+0.007}_{-0.008}$ and only an upper limit on the flash amplitude, $\xflash<0.28$ at 95\% C.L. By contrast, PR4 gives $\taureio=0.060^{+0.005}_{-0.006}$ and a marginalized constraint $\xflash=0.24^{+0.13}_{-0.07}$.

The corresponding shifts in the parameters relevant for the BAO comparison are modest. In the flash model, $\Omega_m$ shifts from $0.314\pm0.007$ in PR3 to $0.311\pm0.007$ in PR4, while $r_dh$ shifts from $99.2\pm0.9\,{\rm Mpc}$ to $99.7\pm0.9\,{\rm Mpc}$. Thus, although PR4 allows a larger flash amplitude and a slightly larger optical depth, the change in the derived BAO-compressed parameters remains small in this fixed-$z_{\rm flash}$ comparison. 

We repeat this analyses for the second benchmark example, fixed $\zflash=25$, with constraints given in Tab.~\ref{tab:pr3_vs_pr4_lcdm_and_flash_big}. We find similar results, including a 95\%CL upper bound $\xflash<0.18$.

This comparison shows that the inferred flash contribution depends on the large-scale polarization data set. We therefore use PR4 as our benchmark Planck data set, while keeping PR3 as a check of the dependence on the Planck likelihood choice.

%%%%%%%%%%%%%%%%%%%%%%%%%%%%%%%%%%%
%%%%%%%%%%%%%%%%%%%%%%%%%%%%%%%%%%%
%%%%%%%%%%%%%%%%%%%%%%%%%%%%%%%%%%%
%%%%%%%%%%%%%%%%%%%%%%%%%%%%%%%%%%%
%\vspace{1cm}
\section{Discussion and conclusion}
\label{sec:discussion}
%%%%%%%%%%%%%%%%%%%%%%%%%%%%%%%%%%%
%%%%%%%%%%%%%%%%%%%%%%%%%%%%%%%%%%%
%%%%%%%%%%%%%%%%%%%%%%%%%%%%%%%%%%%
%%%%%%%%%%%%%%%%%%%%%%%%%%%%%%%%%%%

In this work, we have presented the first full MCMC analysis of $\Lambda$CDM with flash reionization fit to Planck temperature, polarization, and lensing data. Previous studies considered fixed benchmark flash histories and showed that an early ionization episode can increase the optical depth while modifying the low-$\ell$ polarization spectrum. Our goal was to test whether this picture survives once the flash redshift, amplitude, duration, late-time reionization history, and cosmological parameters are varied simultaneously.

The main result is that Planck PR4 allows a modest flash contribution, but the marginalized total optical depth remains close to the standard tanh result. In the flash model with all parameters varied, Fig.~\ref{fig:flagship}, we find $\tau=0.061\pm0.007$, compared with $\tau=0.059\pm0.006$ for the standard tanh parametrization. Thus, although the CMB allows an early ionization component, the fully marginalized analysis does not produce the large increase in $\tau$ suggested by the benchmark high-$\tau$ histories.

At the same time, the minimum-$\chi^2$ sampled point reaches $\tau\simeq0.068$, showing that somewhat higher-$\tau$ flash histories can be accommodated by Planck. The difference between this point and the marginalized posterior is what makes the interpretation of the full Bayesian result nontrivial.

The full MCMC analysis also shows why this result is more subtle than a comparison of fixed histories. In the flash model, $\tau$ is a derived quantity, and uniform priors on the flash parameters induce a non-uniform implicit prior on the total optical depth. We quantified this prior explicitly and found that it has substantial support for $\tau>0.06$, so higher-$\tau$ histories are not excluded a priori by the chosen prior. At the same time, the all-free posterior contains directions that are only weakly constrained by the data, especially in $z_{\rm flash}$ and $t_{\rm rise}$. In this situation, marginalized constraints can depend on how the likelihood is integrated over weakly constrained regions of parameter space and on the prior volume assigned to those regions. Thus, the absence of a large marginalized shift in $\tau$ should be understood as a result of the full marginalized inference, conditioned on the chosen parametrization and priors, rather than as a statement about any single fixed flash history. A cleaner separation of likelihood effects, prior-volume effects, and parametrization dependence would require additional analyses, such as alternative priors, alternative parametrizations, or profile-likelihood studies.

The fixed-redshift analyses make this point more concrete. For the benchmark redshifts $z_{\rm flash}=20$ and $z_{\rm flash}=25$, Planck PR4 gives nonzero flash amplitudes, but these amplitudes are below the benchmark value $\xflash=0.5$ used in previous fixed-history studies. The corresponding total optical depths, $\tau=0.060^{+0.005}_{-0.006}$ and $\tau=0.062\pm 0.006$, remain close to the standard tanh constraint rather than reaching $\tau\simeq 0.09$. This result is similar to that of Sierra et al., Ref.~\cite{Sierra:2026vtz}, in the context of modified reionzation from black hole evaporation, which also found only a modest boost to $\tau$, thus suggesting that these high-redshift scenarios may not be a way out of the CMB-BAO tension. Our work illustrates that a nonzero flash contribution should not be identified directly with a large increase in the total optical depth. In the all-free analysis, the optical-depth budget is redistributed between early and late scattering: the posterior for $\tau_{\rm early}$ has support at zero and a modest second mode around $\tau_{\rm early}\sim 0.012$, while the late contribution remains close to $\tau_{\rm late}\sim 0.05$. Thus, the CMB can allow an early scattering contribution without requiring a substantially larger total $\tau$.

The comparison between Planck PR3 and PR4 shows that the inferred flash contribution is sensitive to the large-scale polarization data set. This is consistent with previous work showing that PR4 is particularly important for analyses of flexible or non-standard reionization histories \cite{Tristram2024,Ilic:2025idl}. PR4 is more permissive of a flash contribution than PR3, but the corresponding shifts in $\tau$, $\Omega_m$, and $r_dh$ remain small.

This analysis also illustrates the difficulty of constraining a detailed reionization history through a single integrated quantity such as $\tau$. Although the low-$\ell$ polarization spectrum is sensitive to the redshift distribution of scattering, Planck data alone do not sharply constrain all directions in the flash parameter space. This is expected for flexible reionization histories and is consistent with previous studies emphasizing the role of reionization priors and large-scale polarization data \cite{MilleaBouchet2018,Ilic:2025idl,Wang:2026bsq}. Future work should therefore explore alternative parametrizations and prior choices, profile-likelihood approaches, and combinations with external probes of reionization, such as Ly$\alpha$ constraints, kSZ measurements, and high-redshift galaxy or quasar data. Such analyses will be essential for determining how robust the flash-reionization interpretation is beyond the specific parametrization and priors adopted here. 

Future work could also perform a more comprehensive study of constraints from varied datasets, including CMB datasets such as ACT \cite{ACT:2025fju} and SPT \cite{SPT-3G:2025bzu}, LSS datasets such as DESI BAO \cite{DESIDR2} and DESI full-shape \cite{DESI:2024jxi}, and CMB-LSS cross correlation. Following Ref.~\cite{Sullivan:2026tas}, it would be interesting to  see how constraints on flash reionization change when low-$\ell$ EE CMB data is removed but LSS data is included. We leave this and other interesting directions to future work.

\acknowledgments

The authors thank Lloyd Knox for helpful discussions. P. Upadhyay is supported by a Canada Graduate Research Scholarship – Master's from the Natural Sciences and Engineering Research Council of Canada (NSERC). E.M. is supported in part by a Discovery Grant from the Natural Sciences and Engineering Research Council of Canada. Kavli IPMU is supported by the World Premier International Research Center Initiative (WPI), MEXT, Japan.

\appendix

%%%%%%%%%%%%%%%%%%%%%%%%%%%%%%%%%%%%%%%%%%%%%%%%%%%%%%
%%%%%%%%%%%%%%%%%%%%%%%%%%%%%%%%%%%%%%%%%%%%%%%%%%%%%%
\section{Supplementary Tables}
%%%%%%%%%%%%%%%%%%%%%%%%%%%%%%%%%%%%%%%%%%%%%%%%%%%%%%
%%%%%%%%%%%%%%%%%%%%%%%%%%%%%%%%%%%%%%%%%%%%%%%%%%%%%%

\begin{table*}[h!]
\centering
Constraints from Planck PR4\\
\begin{tabular}{|l|c|}
\hline\hline
Parameter & PR4 flash \\
\hline\hline
{\boldmath$\ln(10^{10}A_\mathrm{s})$}
& $3.049\;(3.053)\pm0.013$ \\

{\boldmath$n_\mathrm{s}$}
& $0.967\;(0.969)\pm0.004$ \\

{\boldmath$\Omega_\mathrm{b}h^2$}
& $0.02223\;(0.02234)\pm0.00014$ \\

{\boldmath$\Omega_\mathrm{c}h^2$}
& $0.1189\;(0.1184)\pm0.0012$ \\

{\boldmath$z_\mathrm{re}$}
& $7.2\;(7.8)^{+1.1}_{-0.9}$ \\

{\boldmath$z_\mathrm{flash}$}
& $<37.4\;(34.7)$ \\

{\boldmath$x_\mathrm{flash}$}
& $<0.83\;(0.14)$ \\

{\boldmath$t_\mathrm{rise}\,[\mathrm{Myr}]$}
& $<90.5\;(9.5)$ \\

{\boldmath$100\theta_\mathrm{s}$}
& $1.04181\;(1.04177)\pm0.00025$ \\
\hline

{$\tau_\mathrm{reio}$}
& $0.061\;(0.067)\pm0.007$ \\

$H_0\,[\mathrm{km\,s^{-1}\,Mpc^{-1}}]$
& $67.6\;(67.9)\pm0.5$ \\

$\Omega_\mathrm{m}$
& $0.310\;(0.307)\pm0.007$ \\

$\sigma_8$
& $0.810\;(0.810)\pm0.005$ \\

$S_8$
& $0.824\;(0.819)\pm0.012$ \\

$r_\mathrm{d}h\,[\mathrm{Mpc}]$
& $99.7\;(100.1)\pm0.9$ \\
\hline
\end{tabular}
\caption{Constraint on $\Lambda$CDM with flash reionization from Planck PR4 data. Sampled parameters are shown in bold. Parentheses give best sampled posterior point. Upper bounds are given at 95\% CL. \label{tab:theta_s_flagship_full}} 
\end{table*}

\begin{table*}
\centering
\begingroup
\setlength{\tabcolsep}{3pt}
\scriptsize
Comparison of constraints from Planck PR3 and PR4.
\begin{tabular}{|l|c|c|c|c|c|c|}
\hline\hline
Parameter & PR3 tanh & PR4 tanh & PR3 flash $z=20$ & PR4 flash $z=20$ & PR3 flash $z=25$ & PR4 flash $z=25$ \\
\hline\hline
{$z_\mathrm{flash}$} & $-$ & $-$ & $20\;\mathrm{(fixed)}$ & $20\;\mathrm{(fixed)}$ & $25\;\mathrm{(fixed)}$ & $25\;\mathrm{(fixed)}$ \\
{$x_\mathrm{flash}$} & $-$ & $-$ & $< 0.28\;(0.08)$ & $0.24\;(0.21)^{+0.13}_{-0.07}$ & $< 0.18\;(0.08)$ & $0.15\;(0.15)^{+0.07}_{-0.05}$ \\
{$t_\mathrm{rise}\,[\mathrm{Myr}]$} & $-$ & $-$ & $30\;\mathrm{(fixed)}$ & $30\;\mathrm{(fixed)}$ & $30\;\mathrm{(fixed)}$ & $30\;\mathrm{(fixed)}$ \\
{$\tau_\mathrm{reio}$} & $0.054\;(0.044)\pm 0.007$ & $0.059\;(0.061)\pm 0.006$ & $0.057\;(0.053)^{+0.007}_{-0.008}$ & $0.060\;(0.064)^{+0.005}_{-0.006}$ & $0.057\;(0.056)\pm 0.008$ & $0.062\;(0.062)\pm 0.006$ \\
$\Omega_\mathrm{m}$ & $0.315\;(0.317)\pm 0.007$ & $0.311\;(0.312)\pm 0.007$ & $0.314\;(0.314)\pm 0.007$ & $0.311\;(0.305)\pm 0.007$ & $0.314\;(0.312)\pm 0.007$ & $0.310\;(0.310)\pm 0.007$ \\
$r_\mathrm{d}h\,[\mathrm{Mpc}]$ & $99.1\;(98.9)\pm 0.9$ & $99.6\;(99.7)\pm 0.9$ & $99.2\;(99.2)\pm 0.9$ & $99.7\;(100.4)\pm 0.9$ & $99.3\;(99.5)\pm 0.9$ & $99.8\;(99.8)\pm 0.9$ \\
\hline
\end{tabular}
\endgroup
\caption{PR3 and PR4 constraints in tanh and flash reionization. Flash columns compare both fixed redshifts, $z_\mathrm{flash}=20,25$, at $t_\mathrm{rise}=30\,\mathrm{Myr}$.\label{tab:pr3_vs_pr4_lcdm_and_flash_big}}
\end{table*}

\begin{table*}
\centering
Comparison of constraints from Planck PR4 with and without lensing.

\begin{tabular}{|l|c|c|}
\hline\hline
Parameter & PR4 without lensing & PR4 with lensing \\
\hline\hline
{$z_\mathrm{flash}$} & $20\;\mathrm{(fixed)}$ & $20\;\mathrm{(fixed)}$ \\
{$x_\mathrm{flash}$} & $0.24\;(0.27)^{+0.13}_{-0.07}$ & $0.24\;(0.21)^{+0.13}_{-0.07}$ \\
{$t_\mathrm{rise}\,[\mathrm{Myr}]$} & $30\;\mathrm{(fixed)}$ & $30\;\mathrm{(fixed)}$ \\
{$\tau_\mathrm{reio}$} & $0.059\;(0.065)^{+0.005}_{-0.006}$ & $0.060\;(0.064)^{+0.005}_{-0.006}$ \\
$\Omega_\mathrm{m}$ & $0.309\;(0.306)\pm 0.008$ & $0.311\;(0.305)\pm 0.007$ \\
$r_\mathrm{d}h\,[\mathrm{Mpc}]$ & $99.9\;(100.3)\pm 1.0$ & $99.7\;(100.4)\pm 0.9$ \\
\hline
\end{tabular}
\caption{Impact of Planck PR4 lensing on flash constraints at
$z_\mathrm{flash}=20$ and $t_\mathrm{rise}=30\,\mathrm{Myr}$.}
\label{tab:lensing_fixed_trise_zflash}
\end{table*}

\begin{table*}
\centering
Constraints on flash reionization at varied fixed $t_{\rm rise}$.\\
\begin{tabular}{|l|c|c|}
\hline\hline
Parameter & $t_\mathrm{rise}=10\,\mathrm{Myr}$ & $t_\mathrm{rise}=30\,\mathrm{Myr}$ \\
\hline\hline
{$z_\mathrm{flash}$} & $> 7.4\;(35.7)$ & $< 37.8\;(32.6)$ \\
{$x_\mathrm{flash}$} & $0.28\;(0.13)^{+0.06}_{-0.26}$ & $< 0.78\;(0.10)$ \\
{$t_\mathrm{rise}\,[\mathrm{Myr}]$} & $10\;\mathrm{(fixed)}$ & $30\;\mathrm{(fixed)}$ \\
{$\tau_\mathrm{reio}$} & $0.061\;(0.062)\pm 0.006$ & $0.061\;(0.067)\pm 0.006$ \\
$\Omega_\mathrm{m}$ & $0.310\;(0.311)\pm 0.007$ & $0.310\;(0.304)\pm 0.007$ \\
$r_\mathrm{d}h\,[\mathrm{Mpc}]$ & $99.7\;(99.6)\pm 0.9$ & $99.7\;(100.6)\pm 0.9$ \\
\hline
\end{tabular}
\caption{PR4 flash constraints for fixed
$t_\mathrm{rise}=10$ and $30\,\mathrm{Myr}$, with
$z_\mathrm{flash}$, $\xflash$, and standard $\Lambda$CDM parameters sampled.}
\label{tab:varied_fixed_trise}
\end{table*}

\bibliography{refs}

%merlin.mbs apsrev4-1.bst 2010-07-25 4.21a (PWD, AO, DPC) hacked
%Control: key (0)
%Control: author (8) initials jnrlst
%Control: editor formatted (1) identically to author
%Control: production of article title (-1) disabled
%Control: page (0) single
%Control: year (1) truncated
%Control: production of eprint (0) enabled
\begin{thebibliography}{41}%
\makeatletter
\providecommand \@ifxundefined [1]{%
 \@ifx{#1\undefined}
}%
\providecommand \@ifnum [1]{%
 \ifnum #1\expandafter \@firstoftwo
 \else \expandafter \@secondoftwo
 \fi
}%
\providecommand \@ifx [1]{%
 \ifx #1\expandafter \@firstoftwo
 \else \expandafter \@secondoftwo
 \fi
}%
\providecommand \natexlab [1]{#1}%
\providecommand \enquote  [1]{``#1''}%
\providecommand \bibnamefont  [1]{#1}%
\providecommand \bibfnamefont [1]{#1}%
\providecommand \citenamefont [1]{#1}%
\providecommand \href@noop [0]{\@secondoftwo}%
\providecommand \href [0]{\begingroup \@sanitize@url \@href}%
\providecommand \@href[1]{\@@startlink{#1}\@@href}%
\providecommand \@@href[1]{\endgroup#1\@@endlink}%
\providecommand \@sanitize@url [0]{\catcode `\\12\catcode `\$12\catcode `\&12\catcode `\#12\catcode `\^12\catcode `\_12\catcode `\%12\relax}%
\providecommand \@@startlink[1]{}%
\providecommand \@@endlink[0]{}%
\providecommand \url  [0]{\begingroup\@sanitize@url \@url }%
\providecommand \@url [1]{\endgroup\@href {#1}{\urlprefix }}%
\providecommand \urlprefix  [0]{URL }%
\providecommand \Eprint [0]{\href }%
\providecommand \doibase [0]{http://dx.doi.org/}%
\providecommand \selectlanguage [0]{\@gobble}%
\providecommand \bibinfo  [0]{\@secondoftwo}%
\providecommand \bibfield  [0]{\@secondoftwo}%
\providecommand \translation [1]{[#1]}%
\providecommand \BibitemOpen [0]{}%
\providecommand \bibitemStop [0]{}%
\providecommand \bibitemNoStop [0]{.\EOS\space}%
\providecommand \EOS [0]{\spacefactor3000\relax}%
\providecommand \BibitemShut  [1]{\csname bibitem#1\endcsname}%
\let\auto@bib@innerbib\@empty
%</preamble>
\bibitem [{\citenamefont {Abdul~Karim}\ \emph {et~al.}(2025)\citenamefont {Abdul~Karim} \emph {et~al.}}]{DESIDR2}%
  \BibitemOpen
  \bibfield  {author} {\bibinfo {author} {\bibfnamefont {M.}~\bibnamefont {Abdul~Karim}} \emph {et~al.} (\bibinfo {collaboration} {DESI}),\ }\href@noop {} {\  (\bibinfo {year} {2025})},\ \Eprint {http://arxiv.org/abs/2503.14738} {arXiv:2503.14738 [astro-ph.CO]} \BibitemShut {NoStop}%
\bibitem [{\citenamefont {Camphuis}\ \emph {et~al.}(2025)\citenamefont {Camphuis} \emph {et~al.}}]{SPT-3G:2025bzu}%
  \BibitemOpen
  \bibfield  {author} {\bibinfo {author} {\bibfnamefont {E.}~\bibnamefont {Camphuis}} \emph {et~al.} (\bibinfo {collaboration} {SPT-3G}),\ }\href@noop {} {\  (\bibinfo {year} {2025})},\ \Eprint {http://arxiv.org/abs/2506.20707} {arXiv:2506.20707 [astro-ph.CO]} \BibitemShut {NoStop}%
\bibitem [{\citenamefont {Ferreira}\ \emph {et~al.}(2026)\citenamefont {Ferreira}, \citenamefont {McDonough}, \citenamefont {Balkenhol}, \citenamefont {Kallosh}, \citenamefont {Knox},\ and\ \citenamefont {Linde}}]{Ferreira:2025lrd}%
  \BibitemOpen
  \bibfield  {author} {\bibinfo {author} {\bibfnamefont {E.~G.~M.}\ \bibnamefont {Ferreira}}, \bibinfo {author} {\bibfnamefont {E.}~\bibnamefont {McDonough}}, \bibinfo {author} {\bibfnamefont {L.}~\bibnamefont {Balkenhol}}, \bibinfo {author} {\bibfnamefont {R.}~\bibnamefont {Kallosh}}, \bibinfo {author} {\bibfnamefont {L.}~\bibnamefont {Knox}}, \ and\ \bibinfo {author} {\bibfnamefont {A.}~\bibnamefont {Linde}},\ }\href {\doibase 10.1103/lq71-b84v} {\bibfield  {journal} {\bibinfo  {journal} {Phys. Rev. D}\ }\textbf {\bibinfo {volume} {113}},\ \bibinfo {pages} {043524} (\bibinfo {year} {2026})},\ \Eprint {http://arxiv.org/abs/2507.12459} {arXiv:2507.12459 [astro-ph.CO]} \BibitemShut {NoStop}%
\bibitem [{\citenamefont {Ye}\ and\ \citenamefont {Lin}(2025)}]{Ye:2025ark}%
  \BibitemOpen
  \bibfield  {author} {\bibinfo {author} {\bibfnamefont {G.}~\bibnamefont {Ye}}\ and\ \bibinfo {author} {\bibfnamefont {S.-J.}\ \bibnamefont {Lin}},\ }\href@noop {} {\  (\bibinfo {year} {2025})},\ \Eprint {http://arxiv.org/abs/2505.02207} {arXiv:2505.02207 [astro-ph.CO]} \BibitemShut {NoStop}%
\bibitem [{\citenamefont {Jhaveri}\ \emph {et~al.}(2025)\citenamefont {Jhaveri}, \citenamefont {Karwal},\ and\ \citenamefont {Hu}}]{Jhaveri:2025neg}%
  \BibitemOpen
  \bibfield  {author} {\bibinfo {author} {\bibfnamefont {T.}~\bibnamefont {Jhaveri}}, \bibinfo {author} {\bibfnamefont {T.}~\bibnamefont {Karwal}}, \ and\ \bibinfo {author} {\bibfnamefont {W.}~\bibnamefont {Hu}},\ }\href@noop {} {\  (\bibinfo {year} {2025})},\ \Eprint {http://arxiv.org/abs/2504.21813} {arXiv:2504.21813 [astro-ph.CO]} \BibitemShut {NoStop}%
\bibitem [{\citenamefont {Sailer}\ \emph {et~al.}(2025)\citenamefont {Sailer}, \citenamefont {Farren}, \citenamefont {Ferraro},\ and\ \citenamefont {White}}]{Sailer:2025lxj}%
  \BibitemOpen
  \bibfield  {author} {\bibinfo {author} {\bibfnamefont {N.}~\bibnamefont {Sailer}}, \bibinfo {author} {\bibfnamefont {G.~S.}\ \bibnamefont {Farren}}, \bibinfo {author} {\bibfnamefont {S.}~\bibnamefont {Ferraro}}, \ and\ \bibinfo {author} {\bibfnamefont {M.}~\bibnamefont {White}},\ }\href@noop {} {\  (\bibinfo {year} {2025})},\ \Eprint {http://arxiv.org/abs/2504.16932} {arXiv:2504.16932 [astro-ph.CO]} \BibitemShut {NoStop}%
\bibitem [{\citenamefont {Maiolino}\ \emph {et~al.}(2024)\citenamefont {Maiolino} \emph {et~al.}}]{Maiolino:2023zdu}%
  \BibitemOpen
  \bibfield  {author} {\bibinfo {author} {\bibfnamefont {R.}~\bibnamefont {Maiolino}} \emph {et~al.},\ }\href {\doibase 10.1038/s41586-024-07494-x} {\bibfield  {journal} {\bibinfo  {journal} {Nature}\ }\textbf {\bibinfo {volume} {627}},\ \bibinfo {pages} {59} (\bibinfo {year} {2024})},\ \bibinfo {note} {[Erratum: Nature 630, E2 (2024)]},\ \Eprint {http://arxiv.org/abs/2305.12492} {arXiv:2305.12492 [astro-ph.GA]} \BibitemShut {NoStop}%
\bibitem [{\citenamefont {{Furtak}}\ \emph {et~al.}(2024)\citenamefont {{Furtak}}, \citenamefont {{Labb{\'e}}}, \citenamefont {{Zitrin}}, \citenamefont {{Greene}}, \citenamefont {{Dayal}}, \citenamefont {{Chemerynska}}, \citenamefont {{Kokorev}}, \citenamefont {{Miller}}, \citenamefont {{Goulding}}, \citenamefont {{de Graaff}}, \citenamefont {{Bezanson}}, \citenamefont {{Brammer}}, \citenamefont {{Cutler}}, \citenamefont {{Leja}}, \citenamefont {{Pan}}, \citenamefont {{Price}}, \citenamefont {{Wang}}, \citenamefont {{Weaver}}, \citenamefont {{Whitaker}}, \citenamefont {{Atek}}, \citenamefont {{Bogd{\'a}n}}, \citenamefont {{Charlot}}, \citenamefont {{Curtis-Lake}}, \citenamefont {{van Dokkum}}, \citenamefont {{Endsley}}, \citenamefont {{Feldmann}}, \citenamefont {{Fudamoto}}, \citenamefont {{Fujimoto}}, \citenamefont {{Glazebrook}}, \citenamefont {{Juneau}}, \citenamefont {{Marchesini}}, \citenamefont {{Maseda}}, \citenamefont {{Nelson}}, \citenamefont {{Oesch}}, \citenamefont {{Plat}}, \citenamefont
  {{Setton}}, \citenamefont {{Stark}},\ and\ \citenamefont {{Williams}}}]{2024Natur.628...57F}%
  \BibitemOpen
  \bibfield  {author} {\bibinfo {author} {\bibfnamefont {L.~J.}\ \bibnamefont {{Furtak}}}, \bibinfo {author} {\bibfnamefont {I.}~\bibnamefont {{Labb{\'e}}}}, \bibinfo {author} {\bibfnamefont {A.}~\bibnamefont {{Zitrin}}}, \bibinfo {author} {\bibfnamefont {J.~E.}\ \bibnamefont {{Greene}}}, \bibinfo {author} {\bibfnamefont {P.}~\bibnamefont {{Dayal}}}, \bibinfo {author} {\bibfnamefont {I.}~\bibnamefont {{Chemerynska}}}, \bibinfo {author} {\bibfnamefont {V.}~\bibnamefont {{Kokorev}}}, \bibinfo {author} {\bibfnamefont {T.~B.}\ \bibnamefont {{Miller}}}, \bibinfo {author} {\bibfnamefont {A.~D.}\ \bibnamefont {{Goulding}}}, \bibinfo {author} {\bibfnamefont {A.}~\bibnamefont {{de Graaff}}}, \bibinfo {author} {\bibfnamefont {R.}~\bibnamefont {{Bezanson}}}, \bibinfo {author} {\bibfnamefont {G.~B.}\ \bibnamefont {{Brammer}}}, \bibinfo {author} {\bibfnamefont {S.~E.}\ \bibnamefont {{Cutler}}}, \bibinfo {author} {\bibfnamefont {J.}~\bibnamefont {{Leja}}}, \bibinfo {author} {\bibfnamefont {R.}~\bibnamefont {{Pan}}},
  \bibinfo {author} {\bibfnamefont {S.~H.}\ \bibnamefont {{Price}}}, \bibinfo {author} {\bibfnamefont {B.}~\bibnamefont {{Wang}}}, \bibinfo {author} {\bibfnamefont {J.~R.}\ \bibnamefont {{Weaver}}}, \bibinfo {author} {\bibfnamefont {K.~E.}\ \bibnamefont {{Whitaker}}}, \bibinfo {author} {\bibfnamefont {H.}~\bibnamefont {{Atek}}}, \bibinfo {author} {\bibfnamefont {{\'A}.}~\bibnamefont {{Bogd{\'a}n}}}, \bibinfo {author} {\bibfnamefont {S.}~\bibnamefont {{Charlot}}}, \bibinfo {author} {\bibfnamefont {E.}~\bibnamefont {{Curtis-Lake}}}, \bibinfo {author} {\bibfnamefont {P.}~\bibnamefont {{van Dokkum}}}, \bibinfo {author} {\bibfnamefont {R.}~\bibnamefont {{Endsley}}}, \bibinfo {author} {\bibfnamefont {R.}~\bibnamefont {{Feldmann}}}, \bibinfo {author} {\bibfnamefont {Y.}~\bibnamefont {{Fudamoto}}}, \bibinfo {author} {\bibfnamefont {S.}~\bibnamefont {{Fujimoto}}}, \bibinfo {author} {\bibfnamefont {K.}~\bibnamefont {{Glazebrook}}}, \bibinfo {author} {\bibfnamefont {S.}~\bibnamefont {{Juneau}}}, \bibinfo {author}
  {\bibfnamefont {D.}~\bibnamefont {{Marchesini}}}, \bibinfo {author} {\bibfnamefont {M.~V.}\ \bibnamefont {{Maseda}}}, \bibinfo {author} {\bibfnamefont {E.}~\bibnamefont {{Nelson}}}, \bibinfo {author} {\bibfnamefont {P.~A.}\ \bibnamefont {{Oesch}}}, \bibinfo {author} {\bibfnamefont {A.}~\bibnamefont {{Plat}}}, \bibinfo {author} {\bibfnamefont {D.~J.}\ \bibnamefont {{Setton}}}, \bibinfo {author} {\bibfnamefont {D.~P.}\ \bibnamefont {{Stark}}}, \ and\ \bibinfo {author} {\bibfnamefont {C.~C.}\ \bibnamefont {{Williams}}},\ }\href {\doibase 10.1038/s41586-024-07184-8} {\bibfield  {journal} {\bibinfo  {journal} {\nat}\ }\textbf {\bibinfo {volume} {628}},\ \bibinfo {pages} {57} (\bibinfo {year} {2024})},\ \Eprint {http://arxiv.org/abs/2308.05735} {arXiv:2308.05735 [astro-ph.GA]} \BibitemShut {NoStop}%
\bibitem [{\citenamefont {Tan}(2025{\natexlab{a}})}]{Tan2025}%
  \BibitemOpen
  \bibfield  {author} {\bibinfo {author} {\bibfnamefont {J.~C.}\ \bibnamefont {Tan}},\ }\href {\doibase 10.3847/2041-8213/adf8da} {\bibfield  {journal} {\bibinfo  {journal} {Astrophys. J. Lett.}\ }\textbf {\bibinfo {volume} {989}},\ \bibinfo {pages} {L47} (\bibinfo {year} {2025}{\natexlab{a}})},\ \Eprint {http://arxiv.org/abs/2506.18490} {arXiv:2506.18490 [astro-ph.CO]} \BibitemShut {NoStop}%
\bibitem [{\citenamefont {Tan}\ and\ \citenamefont {Komatsu}(2025{\natexlab{a}})}]{TanKomatsu2025}%
  \BibitemOpen
  \bibfield  {author} {\bibinfo {author} {\bibfnamefont {J.~C.}\ \bibnamefont {Tan}}\ and\ \bibinfo {author} {\bibfnamefont {E.}~\bibnamefont {Komatsu}},\ }\href {https://arxiv.org/abs/2510.19647} {\enquote {\bibinfo {title} {{The Impact of Population III.1 Flash Reionization for CMB Polarization and Thomson Scattering Optical Depth}},}\ } (\bibinfo {year} {2025}{\natexlab{a}}),\ \Eprint {http://arxiv.org/abs/2510.19647} {arXiv:2510.19647 [astro-ph.CO]} \BibitemShut {NoStop}%
\bibitem [{\citenamefont {Aggarwal}\ \emph {et~al.}(2026)\citenamefont {Aggarwal}, \citenamefont {Cain}, \citenamefont {Lopez}, \citenamefont {Trac}, \citenamefont {D'Aloisio}, \citenamefont {Tanedo},\ and\ \citenamefont {Tan}}]{Aggarwal:2026ogm}%
  \BibitemOpen
  \bibfield  {author} {\bibinfo {author} {\bibfnamefont {Y.}~\bibnamefont {Aggarwal}}, \bibinfo {author} {\bibfnamefont {C.}~\bibnamefont {Cain}}, \bibinfo {author} {\bibfnamefont {G.}~\bibnamefont {Lopez}}, \bibinfo {author} {\bibfnamefont {H.}~\bibnamefont {Trac}}, \bibinfo {author} {\bibfnamefont {A.}~\bibnamefont {D'Aloisio}}, \bibinfo {author} {\bibfnamefont {P.}~\bibnamefont {Tanedo}}, \ and\ \bibinfo {author} {\bibfnamefont {J.~C.}\ \bibnamefont {Tan}},\ }\href@noop {} {\  (\bibinfo {year} {2026})},\ \Eprint {http://arxiv.org/abs/2606.19459} {arXiv:2606.19459 [astro-ph.CO]} \BibitemShut {NoStop}%
\bibitem [{\citenamefont {Petkova}\ \emph {et~al.}(2026)\citenamefont {Petkova}, \citenamefont {Tan}, \citenamefont {Singh}, \citenamefont {Cammelli}, \citenamefont {Sanati}, \citenamefont {Keller}, \citenamefont {Monaco},\ and\ \citenamefont {Nandal}}]{Petkova:2026clg}%
  \BibitemOpen
  \bibfield  {author} {\bibinfo {author} {\bibfnamefont {M.~A.}\ \bibnamefont {Petkova}}, \bibinfo {author} {\bibfnamefont {J.~C.}\ \bibnamefont {Tan}}, \bibinfo {author} {\bibfnamefont {J.}~\bibnamefont {Singh}}, \bibinfo {author} {\bibfnamefont {V.}~\bibnamefont {Cammelli}}, \bibinfo {author} {\bibfnamefont {M.}~\bibnamefont {Sanati}}, \bibinfo {author} {\bibfnamefont {B.}~\bibnamefont {Keller}}, \bibinfo {author} {\bibfnamefont {P.}~\bibnamefont {Monaco}}, \ and\ \bibinfo {author} {\bibfnamefont {D.}~\bibnamefont {Nandal}},\ }\href@noop {} {\  (\bibinfo {year} {2026})},\ \Eprint {http://arxiv.org/abs/2605.28777} {arXiv:2605.28777 [astro-ph.GA]} \BibitemShut {NoStop}%
\bibitem [{\citenamefont {Chon}\ \emph {et~al.}(2026)\citenamefont {Chon}, \citenamefont {Hirano}, \citenamefont {Ishiyama}, \citenamefont {Chang},\ and\ \citenamefont {Springel}}]{Chon2026}%
  \BibitemOpen
  \bibfield  {author} {\bibinfo {author} {\bibfnamefont {S.}~\bibnamefont {Chon}}, \bibinfo {author} {\bibfnamefont {S.}~\bibnamefont {Hirano}}, \bibinfo {author} {\bibfnamefont {T.}~\bibnamefont {Ishiyama}}, \bibinfo {author} {\bibfnamefont {S.-J.}\ \bibnamefont {Chang}}, \ and\ \bibinfo {author} {\bibfnamefont {V.}~\bibnamefont {Springel}},\ }\href {\doibase 10.1038/s41586-026-10985-8} {\bibfield  {journal} {\bibinfo  {journal} {Nature}\ }\textbf {\bibinfo {volume} {657}},\ \bibinfo {pages} {621} (\bibinfo {year} {2026})},\ \Eprint {http://arxiv.org/abs/2601.04955} {arXiv:2601.04955 [astro-ph.GA]} \BibitemShut {NoStop}%
\bibitem [{\citenamefont {Cen}(2003)}]{Cen:2002zc}%
  \BibitemOpen
  \bibfield  {author} {\bibinfo {author} {\bibfnamefont {R.}~\bibnamefont {Cen}},\ }\href {\doibase 10.1086/375217} {\bibfield  {journal} {\bibinfo  {journal} {Astrophys. J.}\ }\textbf {\bibinfo {volume} {591}},\ \bibinfo {pages} {12} (\bibinfo {year} {2003})},\ \Eprint {http://arxiv.org/abs/astro-ph/0210473} {arXiv:astro-ph/0210473} \BibitemShut {NoStop}%
\bibitem [{\citenamefont {Wyithe}\ and\ \citenamefont {Loeb}(2003)}]{Wyithe:2002qu}%
  \BibitemOpen
  \bibfield  {author} {\bibinfo {author} {\bibfnamefont {J.~S.~B.}\ \bibnamefont {Wyithe}}\ and\ \bibinfo {author} {\bibfnamefont {A.}~\bibnamefont {Loeb}},\ }\href {\doibase 10.1086/367721} {\bibfield  {journal} {\bibinfo  {journal} {Astrophys. J.}\ }\textbf {\bibinfo {volume} {586}},\ \bibinfo {pages} {693} (\bibinfo {year} {2003})},\ \Eprint {http://arxiv.org/abs/astro-ph/0209056} {arXiv:astro-ph/0209056} \BibitemShut {NoStop}%
\bibitem [{\citenamefont {Holder}\ \emph {et~al.}(2003)\citenamefont {Holder}, \citenamefont {Haiman}, \citenamefont {Kaplinghat},\ and\ \citenamefont {Knox}}]{Holder:2003eb}%
  \BibitemOpen
  \bibfield  {author} {\bibinfo {author} {\bibfnamefont {G.}~\bibnamefont {Holder}}, \bibinfo {author} {\bibfnamefont {Z.}~\bibnamefont {Haiman}}, \bibinfo {author} {\bibfnamefont {M.}~\bibnamefont {Kaplinghat}}, \ and\ \bibinfo {author} {\bibfnamefont {L.}~\bibnamefont {Knox}},\ }\href {\doibase 10.1086/377338} {\bibfield  {journal} {\bibinfo  {journal} {Astrophys. J.}\ }\textbf {\bibinfo {volume} {595}},\ \bibinfo {pages} {13} (\bibinfo {year} {2003})},\ \Eprint {http://arxiv.org/abs/astro-ph/0302404} {arXiv:astro-ph/0302404} \BibitemShut {NoStop}%
\bibitem [{\citenamefont {Naselsky}\ and\ \citenamefont {Chiang}(2004)}]{Naselsky:2003dp}%
  \BibitemOpen
  \bibfield  {author} {\bibinfo {author} {\bibfnamefont {P.}~\bibnamefont {Naselsky}}\ and\ \bibinfo {author} {\bibfnamefont {L.-Y.}\ \bibnamefont {Chiang}},\ }\href {\doibase 10.1111/j.1365-2966.2004.07250.x} {\bibfield  {journal} {\bibinfo  {journal} {Mon. Not. Roy. Astron. Soc.}\ }\textbf {\bibinfo {volume} {347}},\ \bibinfo {pages} {795} (\bibinfo {year} {2004})},\ \Eprint {http://arxiv.org/abs/astro-ph/0302085} {arXiv:astro-ph/0302085} \BibitemShut {NoStop}%
\bibitem [{\citenamefont {Colombo}\ \emph {et~al.}(2005)\citenamefont {Colombo}, \citenamefont {Bernardi}, \citenamefont {Casarini}, \citenamefont {Mainini}, \citenamefont {Bonometto}, \citenamefont {Carretti},\ and\ \citenamefont {Fabbri}}]{Colombo:2004uh}%
  \BibitemOpen
  \bibfield  {author} {\bibinfo {author} {\bibfnamefont {L.~P.~L.}\ \bibnamefont {Colombo}}, \bibinfo {author} {\bibfnamefont {G.}~\bibnamefont {Bernardi}}, \bibinfo {author} {\bibfnamefont {L.}~\bibnamefont {Casarini}}, \bibinfo {author} {\bibfnamefont {R.}~\bibnamefont {Mainini}}, \bibinfo {author} {\bibfnamefont {S.~A.}\ \bibnamefont {Bonometto}}, \bibinfo {author} {\bibfnamefont {E.}~\bibnamefont {Carretti}}, \ and\ \bibinfo {author} {\bibfnamefont {R.}~\bibnamefont {Fabbri}},\ }\href {\doibase 10.1051/0004-6361:20041761} {\bibfield  {journal} {\bibinfo  {journal} {Astron. Astrophys.}\ }\textbf {\bibinfo {volume} {435}},\ \bibinfo {pages} {413} (\bibinfo {year} {2005})},\ \Eprint {http://arxiv.org/abs/astro-ph/0408022} {arXiv:astro-ph/0408022} \BibitemShut {NoStop}%
\bibitem [{\citenamefont {Furlanetto}\ and\ \citenamefont {Loeb}(2005)}]{Furlanetto:2004nt}%
  \BibitemOpen
  \bibfield  {author} {\bibinfo {author} {\bibfnamefont {S.}~\bibnamefont {Furlanetto}}\ and\ \bibinfo {author} {\bibfnamefont {A.}~\bibnamefont {Loeb}},\ }\href {\doibase 10.1086/429080} {\bibfield  {journal} {\bibinfo  {journal} {Astrophys. J.}\ }\textbf {\bibinfo {volume} {634}},\ \bibinfo {pages} {1} (\bibinfo {year} {2005})},\ \Eprint {http://arxiv.org/abs/astro-ph/0409656} {arXiv:astro-ph/0409656} \BibitemShut {NoStop}%
\bibitem [{\citenamefont {Sierra}\ \emph {et~al.}(2026)\citenamefont {Sierra}, \citenamefont {Lynch}, \citenamefont {Knox}, \citenamefont {Poulin}, \citenamefont {Balkenhol},\ and\ \citenamefont {Khalife}}]{Sierra:2026vtz}%
  \BibitemOpen
  \bibfield  {author} {\bibinfo {author} {\bibfnamefont {I.}~\bibnamefont {Sierra}}, \bibinfo {author} {\bibfnamefont {G.~P.}\ \bibnamefont {Lynch}}, \bibinfo {author} {\bibfnamefont {L.}~\bibnamefont {Knox}}, \bibinfo {author} {\bibfnamefont {V.}~\bibnamefont {Poulin}}, \bibinfo {author} {\bibfnamefont {L.}~\bibnamefont {Balkenhol}}, \ and\ \bibinfo {author} {\bibfnamefont {A.~R.}\ \bibnamefont {Khalife}},\ }\href@noop {} {\  (\bibinfo {year} {2026})},\ \Eprint {http://arxiv.org/abs/2608.01919} {arXiv:2608.01919 [astro-ph.CO]} \BibitemShut {NoStop}%
\bibitem [{\citenamefont {Lewis}(2008)}]{Lewis:2008wr}%
  \BibitemOpen
  \bibfield  {author} {\bibinfo {author} {\bibfnamefont {A.}~\bibnamefont {Lewis}},\ }\href {\doibase 10.1103/PhysRevD.78.023002} {\bibfield  {journal} {\bibinfo  {journal} {Phys. Rev. D}\ }\textbf {\bibinfo {volume} {78}},\ \bibinfo {pages} {023002} (\bibinfo {year} {2008})},\ \Eprint {http://arxiv.org/abs/0804.3865} {arXiv:0804.3865 [astro-ph]} \BibitemShut {NoStop}%
\bibitem [{\citenamefont {Blas}\ \emph {et~al.}(2011{\natexlab{a}})\citenamefont {Blas}, \citenamefont {Lesgourgues},\ and\ \citenamefont {Tram}}]{CLASS}%
  \BibitemOpen
  \bibfield  {author} {\bibinfo {author} {\bibfnamefont {D.}~\bibnamefont {Blas}}, \bibinfo {author} {\bibfnamefont {J.}~\bibnamefont {Lesgourgues}}, \ and\ \bibinfo {author} {\bibfnamefont {T.}~\bibnamefont {Tram}},\ }\href {\doibase 10.1088/1475-7516/2011/07/034} {\bibfield  {journal} {\bibinfo  {journal} {JCAP}\ }\textbf {\bibinfo {volume} {2011}},\ \bibinfo {pages} {034} (\bibinfo {year} {2011}{\natexlab{a}})},\ \Eprint {http://arxiv.org/abs/1104.2933} {arXiv:1104.2933 [astro-ph.CO]} \BibitemShut {NoStop}%
\bibitem [{\citenamefont {Liu}\ \emph {et~al.}(2025)\citenamefont {Liu}, \citenamefont {Zhu}, \citenamefont {Hu},\ and\ \citenamefont {Miranda}}]{Liu:2025bss}%
  \BibitemOpen
  \bibfield  {author} {\bibinfo {author} {\bibfnamefont {R.}~\bibnamefont {Liu}}, \bibinfo {author} {\bibfnamefont {Y.}~\bibnamefont {Zhu}}, \bibinfo {author} {\bibfnamefont {W.}~\bibnamefont {Hu}}, \ and\ \bibinfo {author} {\bibfnamefont {V.}~\bibnamefont {Miranda}},\ }\href@noop {} {\  (\bibinfo {year} {2025})},\ \Eprint {http://arxiv.org/abs/2510.14957} {arXiv:2510.14957 [astro-ph.CO]} \BibitemShut {NoStop}%
\bibitem [{\citenamefont {Allali}\ \emph {et~al.}(2025)\citenamefont {Allali}, \citenamefont {Li}, \citenamefont {Singh},\ and\ \citenamefont {Fan}}]{Allali:2025yvp}%
  \BibitemOpen
  \bibfield  {author} {\bibinfo {author} {\bibfnamefont {I.~J.}\ \bibnamefont {Allali}}, \bibinfo {author} {\bibfnamefont {L.}~\bibnamefont {Li}}, \bibinfo {author} {\bibfnamefont {P.}~\bibnamefont {Singh}}, \ and\ \bibinfo {author} {\bibfnamefont {J.}~\bibnamefont {Fan}},\ }\href@noop {} {\  (\bibinfo {year} {2025})},\ \Eprint {http://arxiv.org/abs/2509.09678} {arXiv:2509.09678 [astro-ph.CO]} \BibitemShut {NoStop}%
\bibitem [{\citenamefont {Millea}\ and\ \citenamefont {Bouchet}(2018)}]{MilleaBouchet2018}%
  \BibitemOpen
  \bibfield  {author} {\bibinfo {author} {\bibfnamefont {M.}~\bibnamefont {Millea}}\ and\ \bibinfo {author} {\bibfnamefont {F.}~\bibnamefont {Bouchet}},\ }\href {\doibase 10.1051/0004-6361/201833288} {\bibfield  {journal} {\bibinfo  {journal} {Astron. Astrophys.}\ }\textbf {\bibinfo {volume} {617}},\ \bibinfo {pages} {A96} (\bibinfo {year} {2018})},\ \Eprint {http://arxiv.org/abs/1804.08476} {arXiv:1804.08476 [astro-ph.CO]} \BibitemShut {NoStop}%
\bibitem [{\citenamefont {Ilic}\ \emph {et~al.}(2025)\citenamefont {Ilic} \emph {et~al.}}]{Ilic:2025idl}%
  \BibitemOpen
  \bibfield  {author} {\bibinfo {author} {\bibfnamefont {S.}~\bibnamefont {Ilic}} \emph {et~al.},\ }\href {\doibase 10.1051/0004-6361/202555196} {\bibfield  {journal} {\bibinfo  {journal} {Astron. Astrophys.}\ }\textbf {\bibinfo {volume} {700}},\ \bibinfo {pages} {A26} (\bibinfo {year} {2025})},\ \Eprint {http://arxiv.org/abs/2504.13254} {arXiv:2504.13254 [astro-ph.CO]} \BibitemShut {NoStop}%
\bibitem [{\citenamefont {Wang}\ and\ \citenamefont {Shan}(2026)}]{Wang:2026bsq}%
  \BibitemOpen
  \bibfield  {author} {\bibinfo {author} {\bibfnamefont {Z.}~\bibnamefont {Wang}}\ and\ \bibinfo {author} {\bibfnamefont {H.}~\bibnamefont {Shan}},\ }\href@noop {} {\  (\bibinfo {year} {2026})},\ \Eprint {http://arxiv.org/abs/2606.27903} {arXiv:2606.27903 [astro-ph.CO]} \BibitemShut {NoStop}%
\bibitem [{\citenamefont {Aghanim}\ \emph {et~al.}(2020{\natexlab{a}})\citenamefont {Aghanim} \emph {et~al.}}]{Planck2018Likelihood}%
  \BibitemOpen
  \bibfield  {author} {\bibinfo {author} {\bibfnamefont {N.}~\bibnamefont {Aghanim}} \emph {et~al.} (\bibinfo {collaboration} {Planck}),\ }\href {\doibase 10.1051/0004-6361/201936386} {\bibfield  {journal} {\bibinfo  {journal} {Astron. Astrophys.}\ }\textbf {\bibinfo {volume} {641}},\ \bibinfo {pages} {A5} (\bibinfo {year} {2020}{\natexlab{a}})},\ \Eprint {http://arxiv.org/abs/1907.12875} {arXiv:1907.12875 [astro-ph.CO]} \BibitemShut {NoStop}%
\bibitem [{\citenamefont {Akrami}\ \emph {et~al.}(2020)\citenamefont {Akrami} \emph {et~al.}}]{planck20-57}%
  \BibitemOpen
  \bibfield  {author} {\bibinfo {author} {\bibfnamefont {Y.}~\bibnamefont {Akrami}} \emph {et~al.} (\bibinfo {collaboration} {Planck}),\ }\href {\doibase 10.1051/0004-6361/202038073} {\bibfield  {journal} {\bibinfo  {journal} {Astron. Astrophys.}\ }\textbf {\bibinfo {volume} {643}},\ \bibinfo {pages} {A42} (\bibinfo {year} {2020})},\ \Eprint {http://arxiv.org/abs/2007.04997} {arXiv:2007.04997 [astro-ph.CO]} \BibitemShut {NoStop}%
\bibitem [{\citenamefont {Tristram}\ \emph {et~al.}(2024)\citenamefont {Tristram} \emph {et~al.}}]{Tristram2024}%
  \BibitemOpen
  \bibfield  {author} {\bibinfo {author} {\bibfnamefont {M.}~\bibnamefont {Tristram}} \emph {et~al.},\ }\href {\doibase 10.1051/0004-6361/202348015} {\bibfield  {journal} {\bibinfo  {journal} {Astron. Astrophys.}\ }\textbf {\bibinfo {volume} {682}},\ \bibinfo {pages} {A37} (\bibinfo {year} {2024})},\ \Eprint {http://arxiv.org/abs/2309.10034} {arXiv:2309.10034 [astro-ph.CO]} \BibitemShut {NoStop}%
\bibitem [{\citenamefont {Aghanim}\ \emph {et~al.}(2020{\natexlab{b}})\citenamefont {Aghanim} \emph {et~al.}}]{Planck:2018lbu}%
  \BibitemOpen
  \bibfield  {author} {\bibinfo {author} {\bibfnamefont {N.}~\bibnamefont {Aghanim}} \emph {et~al.} (\bibinfo {collaboration} {Planck}),\ }\href {\doibase 10.1051/0004-6361/201833886} {\bibfield  {journal} {\bibinfo  {journal} {Astron. Astrophys.}\ }\textbf {\bibinfo {volume} {641}},\ \bibinfo {pages} {A8} (\bibinfo {year} {2020}{\natexlab{b}})},\ \Eprint {http://arxiv.org/abs/1807.06210} {arXiv:1807.06210 [astro-ph.CO]} \BibitemShut {NoStop}%
\bibitem [{\citenamefont {Carron}\ \emph {et~al.}(2022)\citenamefont {Carron}, \citenamefont {Mirmelstein},\ and\ \citenamefont {Lewis}}]{Carron:2022eyg}%
  \BibitemOpen
  \bibfield  {author} {\bibinfo {author} {\bibfnamefont {J.}~\bibnamefont {Carron}}, \bibinfo {author} {\bibfnamefont {M.}~\bibnamefont {Mirmelstein}}, \ and\ \bibinfo {author} {\bibfnamefont {A.}~\bibnamefont {Lewis}},\ }\href {\doibase 10.1088/1475-7516/2022/09/039} {\bibfield  {journal} {\bibinfo  {journal} {JCAP}\ }\textbf {\bibinfo {volume} {09}},\ \bibinfo {pages} {039} (\bibinfo {year} {2022})},\ \Eprint {http://arxiv.org/abs/2206.07773} {arXiv:2206.07773 [astro-ph.CO]} \BibitemShut {NoStop}%
\bibitem [{\citenamefont {McDonough}\ and\ \citenamefont {Ferreira}(2025)}]{McDonough:2025lzo}%
  \BibitemOpen
  \bibfield  {author} {\bibinfo {author} {\bibfnamefont {E.}~\bibnamefont {McDonough}}\ and\ \bibinfo {author} {\bibfnamefont {E.~G.~M.}\ \bibnamefont {Ferreira}},\ }\href@noop {} {\  (\bibinfo {year} {2025})},\ \Eprint {http://arxiv.org/abs/2512.05108} {arXiv:2512.05108 [astro-ph.CO]} \BibitemShut {NoStop}%
\bibitem [{\citenamefont {Torrado}\ and\ \citenamefont {Lewis}(2019)}]{torrado_lewis_2019}%
  \BibitemOpen
  \bibfield  {author} {\bibinfo {author} {\bibfnamefont {J.}~\bibnamefont {Torrado}}\ and\ \bibinfo {author} {\bibfnamefont {A.}~\bibnamefont {Lewis}},\ }\href {https://github.com/CobayaSampler/cobaya} {\enquote {\bibinfo {title} {Cobaya},}\ } (\bibinfo {year} {2019})\BibitemShut {NoStop}%
\bibitem [{\citenamefont {Blas}\ \emph {et~al.}(2011{\natexlab{b}})\citenamefont {Blas}, \citenamefont {Lesgourgues},\ and\ \citenamefont {Tram}}]{Diego_Blas_2011}%
  \BibitemOpen
  \bibfield  {author} {\bibinfo {author} {\bibfnamefont {D.}~\bibnamefont {Blas}}, \bibinfo {author} {\bibfnamefont {J.}~\bibnamefont {Lesgourgues}}, \ and\ \bibinfo {author} {\bibfnamefont {T.}~\bibnamefont {Tram}},\ }\href {\doibase 10.1088/1475-7516/2011/07/034} {\bibfield  {journal} {\bibinfo  {journal} {Journal of Cosmology and Astroparticle Physics}\ }\textbf {\bibinfo {volume} {2011}},\ \bibinfo {pages} {034–034} (\bibinfo {year} {2011}{\natexlab{b}})}\BibitemShut {NoStop}%
\bibitem [{\citenamefont {Lewis}(2025)}]{GetDist}%
  \BibitemOpen
  \bibfield  {author} {\bibinfo {author} {\bibfnamefont {A.}~\bibnamefont {Lewis}},\ }\href {\doibase 10.1088/1475-7516/2025/08/025} {\bibfield  {journal} {\bibinfo  {journal} {JCAP}\ }\textbf {\bibinfo {volume} {2025}},\ \bibinfo {pages} {025} (\bibinfo {year} {2025})},\ \Eprint {http://arxiv.org/abs/1910.13970} {arXiv:1910.13970 [astro-ph.IM]} \BibitemShut {NoStop}%
\bibitem [{\citenamefont {Tan}\ and\ \citenamefont {Komatsu}(2025{\natexlab{b}})}]{Tan:2025obi}%
  \BibitemOpen
  \bibfield  {author} {\bibinfo {author} {\bibfnamefont {J.~C.}\ \bibnamefont {Tan}}\ and\ \bibinfo {author} {\bibfnamefont {E.}~\bibnamefont {Komatsu}},\ }\href@noop {} {\  (\bibinfo {year} {2025}{\natexlab{b}})},\ \Eprint {http://arxiv.org/abs/2510.19647} {arXiv:2510.19647 [astro-ph.CO]} \BibitemShut {NoStop}%
\bibitem [{\citenamefont {Tan}(2025{\natexlab{b}})}]{Tan:2025cua}%
  \BibitemOpen
  \bibfield  {author} {\bibinfo {author} {\bibfnamefont {J.~C.}\ \bibnamefont {Tan}},\ }\href {\doibase 10.3847/2041-8213/adf8da} {\bibfield  {journal} {\bibinfo  {journal} {Astrophys. J. Lett.}\ }\textbf {\bibinfo {volume} {989}},\ \bibinfo {pages} {L47} (\bibinfo {year} {2025}{\natexlab{b}})},\ \Eprint {http://arxiv.org/abs/2506.18490} {arXiv:2506.18490 [astro-ph.CO]} \BibitemShut {NoStop}%
\bibitem [{\citenamefont {Louis}\ \emph {et~al.}(2025)\citenamefont {Louis} \emph {et~al.}}]{ACT:2025fju}%
  \BibitemOpen
  \bibfield  {author} {\bibinfo {author} {\bibfnamefont {T.}~\bibnamefont {Louis}} \emph {et~al.} (\bibinfo {collaboration} {ACT}),\ }\href@noop {} {\  (\bibinfo {year} {2025})},\ \Eprint {http://arxiv.org/abs/2503.14452} {arXiv:2503.14452 [astro-ph.CO]} \BibitemShut {NoStop}%
\bibitem [{\citenamefont {Adame}\ \emph {et~al.}(2025)\citenamefont {Adame} \emph {et~al.}}]{DESI:2024jxi}%
  \BibitemOpen
  \bibfield  {author} {\bibinfo {author} {\bibfnamefont {A.~G.}\ \bibnamefont {Adame}} \emph {et~al.} (\bibinfo {collaboration} {DESI}),\ }\href {\doibase 10.1088/1475-7516/2025/09/008} {\bibfield  {journal} {\bibinfo  {journal} {JCAP}\ }\textbf {\bibinfo {volume} {09}},\ \bibinfo {pages} {008} (\bibinfo {year} {2025})},\ \bibinfo {note} {[Erratum: JCAP 02, E02 (2026)]},\ \Eprint {http://arxiv.org/abs/2411.12021} {arXiv:2411.12021 [astro-ph.CO]} \BibitemShut {NoStop}%
\bibitem [{\citenamefont {Sullivan}\ \emph {et~al.}(2026)\citenamefont {Sullivan}, \citenamefont {de~Belsunce},\ and\ \citenamefont {Ivanov}}]{Sullivan:2026tas}%
  \BibitemOpen
  \bibfield  {author} {\bibinfo {author} {\bibfnamefont {J.~M.}\ \bibnamefont {Sullivan}}, \bibinfo {author} {\bibfnamefont {R.}~\bibnamefont {de~Belsunce}}, \ and\ \bibinfo {author} {\bibfnamefont {M.~M.}\ \bibnamefont {Ivanov}},\ }\href@noop {} {\  (\bibinfo {year} {2026})},\ \Eprint {http://arxiv.org/abs/2606.30903} {arXiv:2606.30903 [astro-ph.CO]} \BibitemShut {NoStop}%
\end{thebibliography}%

\end{document}